\documentclass[%
 pre,
rsi,%
 amsmath,amssymb,
 reprint,%
superscriptaddress
]{revtex4-1}

\usepackage{graphicx}% Include figure files
\usepackage{dcolumn}% Align table columns on decimal point
\usepackage{bm}% bold math
\usepackage{color}
\definecolor{cream}{RGB}{222,217,201}
\usepackage{bm}
\usepackage{xpatch}
\usepackage{appendix}
\usepackage[capitalize]{cleveref}
\crefname{appendix}{Appendix-}{Appendices}
\crefname{section}{Sec.\!}{Sections}
\crefname{figure}{Fig.\!\!}{Figures}
\crefrangelabelformat{equation}{(#3#1#4)$-$(#5#2#6)}
\xapptocmd\appendices{%
  \crefalias{section}{appendix}%
}{}{\PatchFailed}

\usepackage[normalem]{ulem}

\let\ref\cref
\usepackage{bbm}
\makeatletter
\def\G_#1{\mathbf{G}_{#1}\@ifnextchar[{\Ebrac}{\relax}}
\def\Ebrac[#1]{#1}
\makeatother

\makeatletter
\def\upd_#1{\mathrm{d} \mathrm{#1}\@ifnextchar[{\Ebrac}{\relax}}
\def\Ebrac[#1]{#1}
\makeatother

\usepackage{tikz}

\definecolor{cadmiumgreen}{rgb}{0.0,0.42,0.24}

\newcommand{\complexi}{\mathbbm{i}}

\begin{document}
% define short cuts !!
\newcommand{\tikzcircle}[2][red,fill=red]{\tikz[baseline=-0.5ex]\draw[#1,radius=#2] (0,0) circle ;}%
\def\bea{\begin{eqnarray}}
\def\eea{\end{eqnarray}}
\def\beq{\begin{equation}}
\def\eeq{\end{equation}}
\def\f{\frac}
\def\k{\kappa}
\def\e{\epsilon}
\def\ve{\varepsilon}
\def\be{\beta}
\def\D{\Delta}
\def\h{\theta}
\def\t{\tau}
\def\a{\alpha}

\def\cDa{{\cal D}[X]}
\def\cD{{\cal D}[x]}
\def\cL{{\cal L}}
\def\cLo{{\cal L}_0}
\def\cLa{{\cal L}_1}

\def\rv{\bf r}
\def\Re{{\rm Re}}
\def\sj{\sum_{j=1}^2}
\def\rk{\rho^{ (k) }}
\def\rek{\rho^{ (1) }}
\def\cek{C^{ (1) }}
\def\rz{\rho^{ (0) }}
\def\rt{\rho^{ (2) }}
\def\rtb{\bar \rho^{ (2) }}
\def\trk{\tilde\rho^{ (k) }}
\def\trek{\tilde\rho^{ (1) }}
\def\trz{\tilde\rho^{ (0) }}
\def\trt{\tilde\rho^{ (2) }}
\def\r{\rho}
\def\tD{\tilde {D}}

\def\s{\sigma}
\def\kb{k_B}
\def\bF{\bar{\cal F}}
\def\F{{\cal F}}
\def\la{\langle}
\def\ra{\rangle}
\def\nn{\nonumber}
\def\up{\uparrow}
\def\dn{\downarrow}
\def\S{\Sigma}
\def\dg{\dagger}
\def\d{\delta}
\def\p{\partial}
\def\l{\lambda}
\def\L{\Lambda}
\def\o{\Omega}
\def\w{\omega}
\def\g{\gamma}

\def\jv{ {\bf j}}
\def\jr{ {\bf j}_r}
\def\jd{ {\bf j}_d}
\def\jdd{ { j}_d}
\def\noi{\noindent}
\def\a{\alpha}
\def\d{\delta}
\def\p{\partial} 
\def\hf{\frac{1}{2}}

\def\la{\langle}
\def\ra{\rangle}
\def\e{\epsilon}
\def\n{\eta}
\def\g{\gamma}
\def\break#1{\pagebreak \vspace*{#1}}
\def\hf{\frac{1}{2}}

\def\rv{{\bf r}}
\def\pc{\phi_c}
\def\rb{\bar\rho}
%% Definitions done-----------------------

\title{Structure-dynamics relationship in ratcheted colloids: Resonance melting, dislocations, and defect clusters}
\author{Shubhendu Shekhar Khali}
\affiliation{Department of Physical Sciences, Indian Institute of Science Education and Research Mohali, Sector 81, S.A.S. Nagar, Manauli-140306,Punjab,India}
\author{Dipanjan  Chakraborty}
\email{chakraborty@iisermohali.ac.in}
\affiliation{Department of Physical Sciences, Indian Institute of Science Education and Research Mohali, Sector 81, S.A.S. Nagar, Manauli-140306,Punjab,India}
\author{Debasish Chaudhuri}
\email{debc@iopb.res.in}
\affiliation{Institute of Physics, Sachivalaya Marg, Bhubaneswar 751005, India.} 
\affiliation{Homi Bhaba National Institute, Anushaktigar, Mumbai 400094, India.}

\date{\today}

\begin{abstract}
We consider a two dimensional colloidal dispersion of soft-core particles driven by a one dimensional stochastic flashing ratchet that induces a time averaged directed particle current through the system. It undergoes a  non-equilibrium melting transition as the directed current approaches a maximum associated with a resonance of the ratcheting frequency with the relaxation frequency of the system. We use extensive molecular dynamics simulations to present a detailed phase diagram in the ratcheting rate- mean density plane. With the help of numerically calculated structure factor, solid and hexatic order parameters, and pair correlation functions, we show that the non-equilibrium melting is a continuous transition from a quasi-long ranged ordered solid to a hexatic phase. The transition is mediated by the unbinding of  dislocations, and formation of compact and string-like defect clusters.
\end{abstract}

\maketitle

%%%MAIN TEXT%%%%
\section{Introduction}
\label{sec:introduction}
 
A class of non-equilibrium driven systems called pump models are
particularly intriguing due to their following property. They involve
periodic forces, in time and space, that vanish under spatio-temporal
averaging but still drives an overall directed
current~\cite{Julicher1997a, Astumian2002, Hanggi2009,Reimann2002,
  Brouwer1998, Citro2003, Jain2007, Chaudhuri2011, Chaudhuri2015,
  Chaudhuri2015f}.  This is achieved via the breaking of time-reversal
symmetry through, e.g., a phase lag between spatially non-local
drives~\cite{Brouwer1998,Jain2007, Chaudhuri2011}, or breaking of
space inversion symmetry of the external potential
profile~\cite{Julicher1997a,Reimann2002, Astumian2002,
  Hanggi2009}. 
  Most of the biological processes generating directed motion involve reaction cycles and utilize some variant of this principle. 
  Natural examples involve ion-pumps, e.g., the Na$^+$,
K$^+$-ATPase pumps, and molecular motors~\cite{Gadsby2009}, e.g.,
Kinesin or myosin moving on polymeric tracks of microtubules or
F-actins, respectively~\cite{Reimann2002}.  
The flashing ratchet model has been used to describe molecular motor locomotion~\cite{Julicher1997a}.  
In experiments on colloids, ratcheting could be generated using
optical~\cite{Faucheux1995,Lopez2008},
magnetic~\cite{Tierno2010,Tierno2012} or electrical
fields~\cite{Rousselet1994, Leibler1994, Marquet2002}.  
Most of the studies on pump models focused on systems of non-interacting
particles, restricted to one dimension, with a few exceptions that
analyzed the impact of interaction on molecular
motors~\cite{Derenyi1995, Derenyi1996}, collective properties of
particle pumps~\cite{Jain2007, Marathe2008,
  Chaudhuri2011,Chaudhuri2015,Chaudhuri2015f}, and in ratchet
models~\cite{Savelev2004, Pototsky2010, Savelev2003, Hanggi2009}.  

In a recent study, we used an asymmetric periodic potential that switches between an {\em on} and {\em off} state in a stochastic manner to drive a directed current of particles in a two dimensional (2d) dispersion of sterically stabilized colloids~\cite{Chakraborty2014}, focusing on the frequency and density dependence of the ratcheted current.   
With the change in the rate of ratcheting, the time- averaged directed current carried by the colloids show a resonance with the system's relaxation frequency~\cite{Chakraborty2014}.  
The current shows a non-monotonic dependence on density as well. This change in the dynamical properties, as we show in this paper, is closely related to the associated structural changes, e.g., the solid melts near the resonance frequency.

In the limit of extremely high switching frequency, higher than the inherent relaxation time of the colloids, the system can only respond to essentially a time- averaged  potential profile. 
In addition, if one considers the limit of vanishing asymmetry in the  potential profile, the scenario becomes equivalent to that of the re-entrant laser induced melting transition (RLIM)~\cite{Chowdhury1985, Wei1998, Frey1999, Chaudhuri2006}, in which a high- density colloidal liquid undergoes solidification followed by melting, as the strength of a commensurate  external periodic potential is increased. 
This is an equilibrium phase transition of the Kosterlitz-Thouless type~\cite{Frey1999, Chaudhuri2006}, and is described in terms of unbinding of a specific type of dislocations, allowed by the potential
profile.
  
In this paper we consider an asymmetric ratcheting of soft-core particles, and investigate structural transitions associated with the change in dynamical behavior of the system, observed in terms of its current carrying capacity.  
Using a large scale molecular dynamics simulation, we obtain the phase diagram in the density- ratcheting rate plane, showing
melting from a solid to hexatic phase. 
We find a re-entrant solid- hexatic- solid transition with changing ratcheting frequency. 
The transitions are associated with a non-monotonic variation of the mean directed current.  
As we demonstrate in detail, the non-equilibrium melting is a continuous transition from a quasi- long ranged ordered (QLRO) solid to a hexatic phase, and is mediated by the formation of topological defects. 
The dominant defect types generated at the solid melting are dislocations, and compact or string-like defect clusters.

%%%%%%%%%%%%
\begin{figure*}[!t]
\centering
\includegraphics[width=\linewidth]{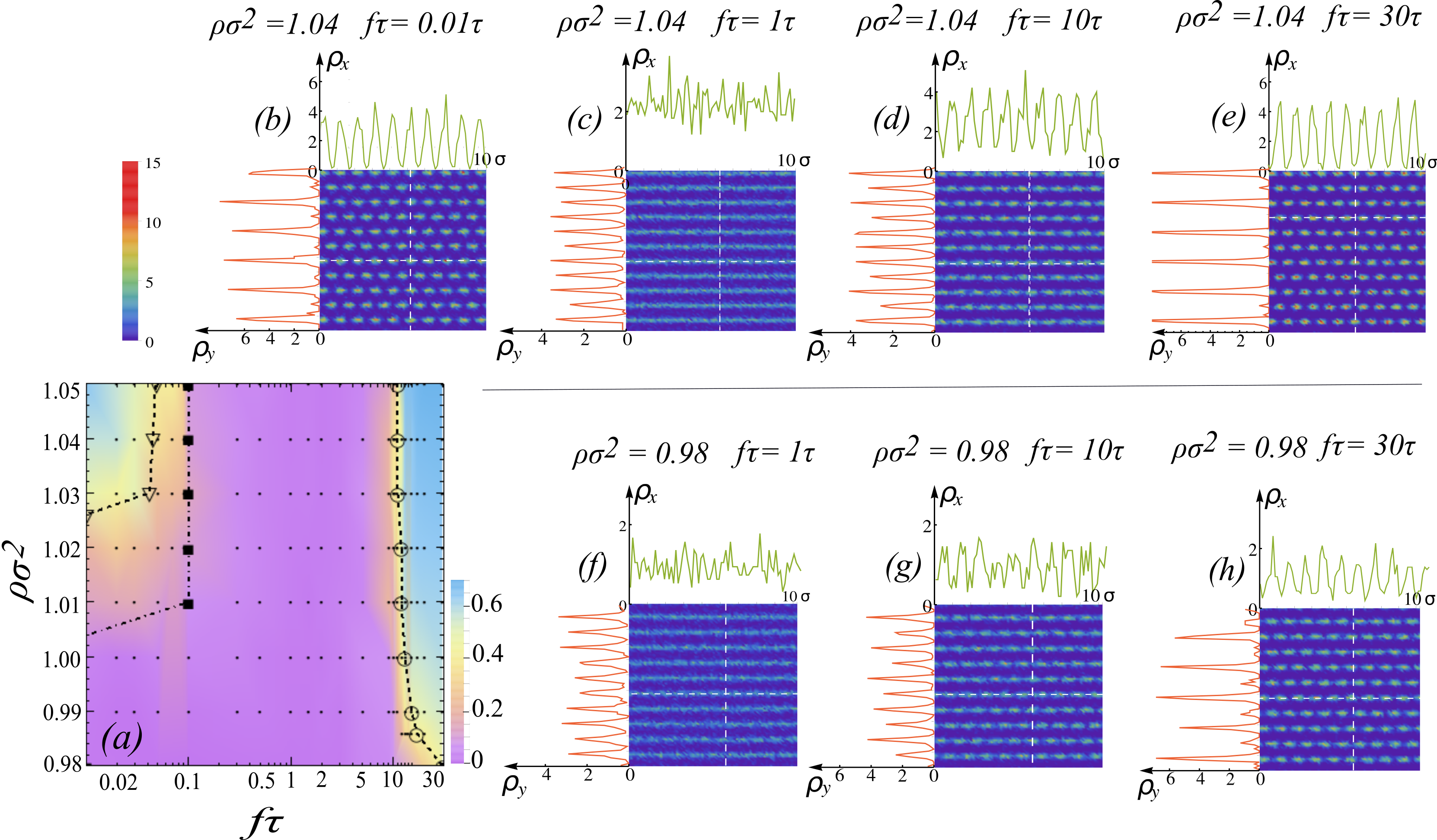}
\caption{
($a$) Phase diagram in the density- frequency plane. The color code indicates  the values of the solid order parameter $\psi_G$ of the
stochastically ratcheted 2d colloidal suspension at different density and ratcheting frequency. Sky-blue denotes high solid order. 
The dashed lines through open symbols ($\triangledown$, $\circ$) show
the boundaries of instability of the solid phase. The dash-dotted line through $\blacksquare$  indicates the low driving frequency $f_s$ below which the system can relax to instantaneous external potential profile. 
In the opposite limit of high frequencies, the system feels a time- integrated constant 
confinement analogous to a laser induced freezing and consequently exhibits a high value of the order parameter. In the regime of
intermediate frequencies, the long range order is broken due to the particle current.  ($b$)-($h$)~Plots of time-averaged local density profiles $\rho(x,y)$ over a section of $10 \s \times 10 \s$ area, at different densities and frequencies as indicated in the figures. Their cuts $\rho_y$ measured along the vertical white dashed lines, and $\rho_x$ measured along the horizontal dashed lines are shown in the out-ward projected ordinate and abscissa, respectively.}
\label{fig:phase_diagram}
\end{figure*}

In \cref{sec:model_simulation} we present the model and details of numerical simulations. 
In \cref{sec:results_discussion} we discuss the detailed phase diagram, explaining the properties of the different non-equilibrium phases. 
The associated variation of driven directed current with driving frequency and density is shown in \cref{ssec:current}. 
In this section we establish the relation of changing particle current to the non-equilibrium phase transitions.  
This is followed by a detailed analysis of the melting transitions in terms of the order parameters presented in \cref{ssec:reentrant_transition}. 
In the following three subsections, the phase- transitions are further characterized in terms of the distribution functions of order parameters, correlation functions, and formation of topological defects. 
We finally conclude presenting a discussion and outlook in \cref{sec:outlook}.

\section{Model and Simulation Details}
\label{sec:model_simulation}

We consider a two dimensional system of a repulsively interacting colloidal suspension of $N$ particles in a volume $A=L_x L_y$.
The mean inter-particle separation in this system $a^2=\sqrt{3} \rho/2$ is set by the particle density $\rho = N/A$. 
We assume that the colloids repel each other via a shifted soft-core potential $U(r)=\e \,[(\sigma/r)^{12}-2^{-12}]$ when the inter-particle separation $r< r_c r_c=2 \sigma$, and $U(r)=0$ otherwise. 
The units of energy and length scales are set by $\e$, $\s$ respectively.  
The system evolves under an asymmetric ratchet potential 
$U_{\rm ext}(x,y,t)=V(t) \left[\sin \left(2 \pi y/\lambda \right)+\alpha \sin \left(4 \pi y/\lambda \right) \right]$, where the time-dependent strength $V(t)$ switches between $\e$ and $0$ stochastically with a rate $f$. 
The two sinusoidal terms in the above expression of $U_{\rm ext}$ with $\a=0.2$ maintains the  asymmetric shape of the potential profile. 
When it assumes a triangular lattice structure, the separation between consecutive lattice planes in the system is $a_y=\sqrt{3} a/2$. 
We have chosen the periodicity of the external potential $\lambda = a_y$, commensurate with the mean lattice spacing. 
In the absence of the external potential, the soft core solid is expected to undergo a two stage solid- hexatic- liquid transition~\cite{Kosterlitz1973, Halperin1978, Young1979, Kapfer:2015ca}, with the solid melting point at $\r \s^2 \approx 1.01$. 
In the presence of a time- independent potential profile with $V(t)=U_0$ and $\a=0$, the system undergoes RLIM with increase in $U_0$~\cite{Wei1998,Frey1999,Chaudhuri2006}. 
At $U_0=\e$, the laser induced melting point of the soft-core solid is $\r \s^2 = 0.95$~\cite{Chaudhuri2006}.

We perform molecular dynamics simulations of the system in the presence of an external ratcheting potential using the standard leap-frog algorithm~\cite{Frenkel2002} with a time-step $\d t = 0.001\,\t$ where $\t= \s \sqrt{m/\e}$ is the characteristic time scale. We use $m=1$. 
The temperature of the system is kept constant at $T =1.0 \e/\kb$ using a Langevin thermostat characterized by an isotropic friction $\g = 1/\t$. 
At each step a trial move is performed to switch $V(t)$ between $0$ and $\e$, and accepted with a probability $f \, \d t$.  
In this paper, we present the results for a large system of $N=262144$ particles. 
We discard simulations over initial $10^7$ steps to ensure achievement of steady state, and the analyses are performed collecting data over further $10^7$ steps.

\section{Results and Discussion}
\label{sec:results_discussion}

\subsection{Phase diagram}
\label{ssec:phase_diagram}

In \cref{fig:phase_diagram} we present the detailed phase diagram along with local density profiles of the 2D system of mono-dispersed ratcheted colloids.  
The system displays a solid and a density modulated hexatic phase,  controlled by the dimensionless density $\r \s^2$ and ratcheting rate $f \t$. 
The color codes in \cref{fig:phase_diagram}($a$) denote the values of mean solid order parameter.
The two dashed lines with open symbols signify  the two solid melting
boundaries at small and high frequencies.  
The dash-dotted line with filled squares denotes the inverse of relaxation time-scales $f_s$ of the system at a given density. For ratcheting rates slower than this time-scale, the solid and hexatic order can {\em equilibrate} to the instantaneous external potential and follow its change. 
The details of the calculation of such relaxation times are discussed in \cref{appendix:relaxtion}.  
We characterize the various phases using the structure factor, the pair
correlation function, the solid and hexatic order parameters, and
their distribution functions.  Further  details of such characterization are
presented in later sections.

As is shown in \cref{fig:phase_diagram}($a$), the system remains in a QLRO triangular lattice solid phase at the highest frequencies, if the  ambient density permits. 
 As the frequency decreases, the solid melts into a hexatic phase, below the dashed line through open circles~(\cref{fig:phase_diagram}). 
 As is shown later, the melting is a continuous transition and happens via a proliferation of topological defects, including the unbinding of dislocation pairs. 
 The molten phase is a hexatic displaying a unimodal distribution of local- hexatic order parameter. 
 This excludes any possibility of phase coexistence, indicating a continuous transition. 
As the frequency of the drive is decreased further, below the equilibrium relaxation time, the system starts to follow the time variation of the external potential.  
As a result, the time-averaged properties turn out to be approximately a superposition of the properties of the equilibrium states in the presence ($V(t)=\e$) and absence ($V(t)=0$) of external potential profile.  
At high densities ($\r \s^2 \gtrsim 1.026$) the solid phase is stabilized even at low ratcheting frequencies like $f\t=0.01$. 
The melting boundary of this solid is shown by the open $\triangledown$ and dashed line.

In \cref{fig:phase_diagram}($b$)-($h$) we present the time-averaged local density profiles $\r(x,y)$ at two mean densities $\r \s^2=1.04,\, 0.98$. 
We show the results over local cross- sections of area $10\s \times 10\s$, for better visibility. 
At $\r \s^2=1.04$, the density profile shows triangular lattice structure at $f \t=0.01$~(\cref{fig:phase_diagram}($b$)\,). 
At $f \t=1$ the solid melts into a phase with density modulation along $y$-axis, the
direction of ratcheting drive~(\cref{fig:phase_diagram}($c$)\,). 
The melting is quantified later in terms of vanishingly small solid order. 
At $f \t=10$, a local triangular lattice- like pattern reappears, albeit the solid order
remains small~(\cref{fig:phase_diagram}($d$)\,). 
The system shows a more compact triangular lattice solid at $f\t=30$ associated with appearance of significant solid order~(\cref{fig:phase_diagram}($e$)\,). 
As is turns out, this order decreases with the system size in a power law manner, identifying a QLRO solid~(see \cref{fig:scaling_solid_order_parameter}).  
As we show later, the density modulated phases of the system at $f \t=1$ and $10$ share similar amount of hexatic order~(see \cref{fig:solid_op_freq_density}).

In the top and left axes around the $\r(x,y)$ plots, we show the linear density profiles $\r_x$ (ordinate) and $\r_y$ (abscissa) measured along the horizontal and vertical dashed white lines indicated in the $\r(x,y)$ plots. 
The clean density modulations in $\r_x$ captures the spontaneous emergence of the solid- like order in that direction. 
Due to the shape of the triangular lattice solid, $\r_y$ shows large followed by tiny peaks along the white line perpendicular to the lattice planes, e.g., in \cref{fig:phase_diagram}($b$) and ($e$). 
Note that this feature is not shared by the other local density plots after the solid melts, shown in \cref{fig:phase_diagram}.  
The strong density modulation $\r_y$ in the hexatic phase is induced directly by the external  potential minima.  
This phase lacks density modulation in $\r_x$, as is shown in \cref{fig:phase_diagram}($c$). 
However, as we shall quantify later, it shows significant hexatic orientation order.

Similar characterization at the lower density  $\r \s^2 = 0.98$ are presented in \cref{fig:phase_diagram}($f$)-($h$).  
A clear solid- like triangular lattice order is observed at $f\t=30$~(\cref{fig:phase_diagram}($h$)\,). 
\cref{fig:phase_diagram}($g$) shows local weak triangular lattice like pattern that gets smeared within the hexatic phase. 
As we show later in Sec.-\ref{ssec:reentrant_transition}, this phase  in \cref{fig:phase_diagram}($f$) and ($g$) indeed display a vanishingly small solid order, in the presence of a reasonably large hexatic order. 
The system remains in the molten phase at the  intermediate frequency range, between $0.1 \leq f \tau \leq 10$. 
Note that the densities considered here are relatively large with respect to the equilibrium melting point in the absence of external potential. 
As we demonstrate in the following, the non-equilibrium  melting discussed in this section is associated with an increase in directed particle current carried by the system under the flashing ratchet drive.

\subsection{Direct particle current out of alternating drive}
\label{ssec:current}
The inherent asymmetry of the ratchet potential, added with the stochastic switching drives a time- and space- averaged directed particle current in the system, 
\begin{equation}
  \label{eq:current_def}
\left\langle  j_y
\right\rangle=\frac{1}{\tau_m}\frac{1}{A}\int^{\tau_m}
\upd_t\int^{L_x}\upd_x\int^{L_y}\upd_y ~j_y(x,y,t),
\end{equation}
where, $\tau_m$ is an integral multiple of $1/f$, the mean switching time of the flashing ratchet potential. 
\begin{figure}[!t]
  \centering
  \includegraphics[width=\linewidth]{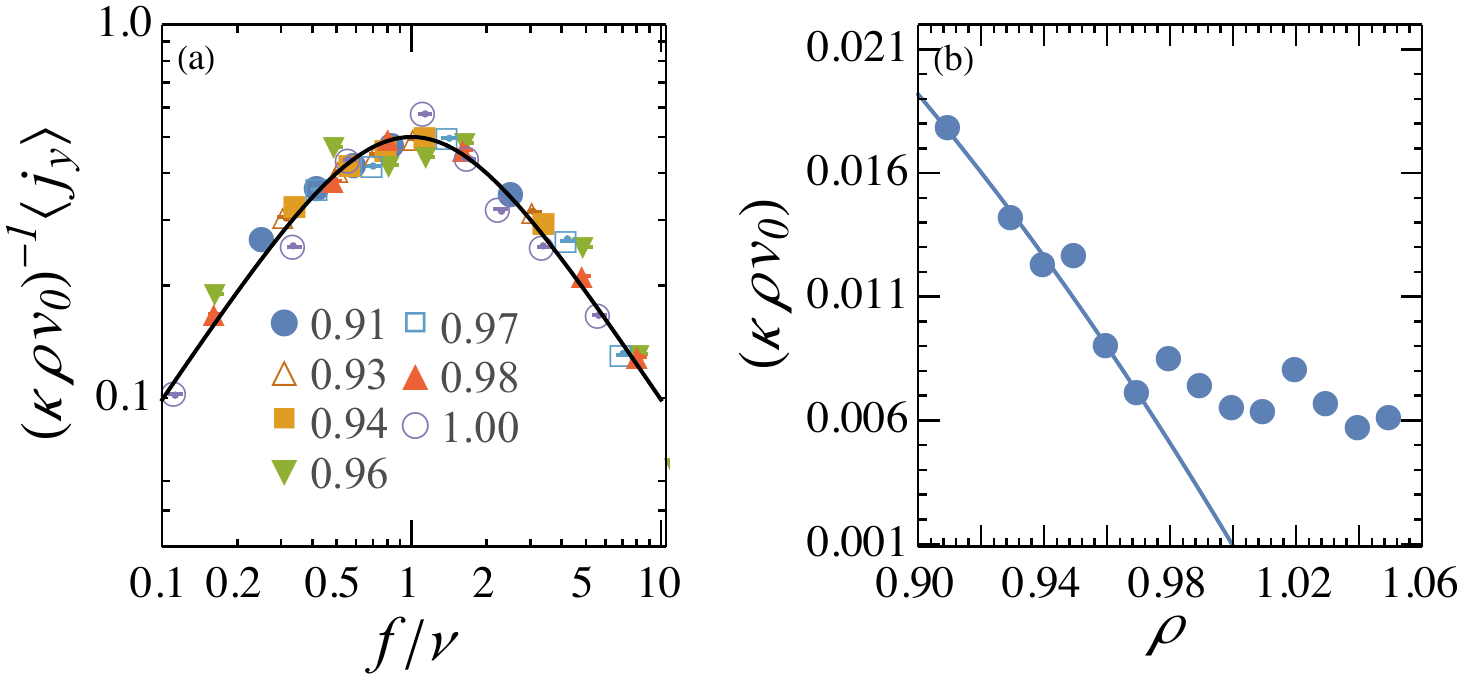}
  \caption{(color online ) {(a)}~Variation of the scaled particle
    current along the direction of the drive as a function $f/\nu$ for
    different densities as indicated in the legend. The solid line is
    plot of the function $(f/\nu)/\left(1+(f/\nu)^2\right)$ as given by \cref{eq:current}.  
    {(b)}~Plot of the current amplitude
    $\rho \kappa v_0$ as a function of density. The solid line is a
    fit to the data using the functional form
    $\kappa D_0 \rho^{3/2}(1-\rho/\r_c)$ with $\kappa D_0$ and $\rho_c$ as fitting 
    parameters with values $\kappa D_0 \approx 0.22$
    and $\rho_c \s^2 \approx 1.004$.  }
  \label{fig:fluxplot}
\end{figure}
The competition between the intrinsic relaxation of the system and
the external drive leads to a resonance in the mean particle current. At low
frequencies the particle current increases linearly with the driving
frequency $f$, achieves a maximum around $f=\nu$,  and beyond
this decays as $f^{-1}$. The behaviour of the current in the whole
frequency range can be captured by the simple ansatz~\cite{Chakraborty2014}
\begin{equation}
  \label{eq:current}
  \left\langle j_y \right\rangle = \kappa \frac{\nu f}{\nu^2+f^2}
  \rho v_0,
\end{equation}
where, $\nu$ is the intrinsic relaxation frequency, $v_0 = \nu \l$
intrinsic velocity, and $\kappa$ is a proportionality constant.

In the high density regime, where, the mean-free path of the particles is small, the diffusive time scale $\tau_D$ to travel the typical distance $\lambda$ is set by $\tau_D=\lambda^2/D(\rho)$. 
Here $D(\rho)$ denotes the density- dependent tagged particle diffusivity, which we assume to decrease linearly with density, 
$D(\rho)=D_0(1-\rho/\rho_c)$~\cite{Lahtinen2001, Falck2004}. 
Here $D_0=\kb T/\gamma$ is the bare diffusivity. The commensurate external
potential ensures that $\lambda^2 \sim 1/\rho$, and consequently the
intrinsic relaxation frequency takes the form $\nu=\rho D(\rho)$. The
intrinsic velocity scale is set by
$v_0=\lambda/\tau_D= \rho^{1/2}D(\rho)$. Substituting for $v_0$ and
$\nu$ in \cref{eq:current}, the density and frequency dependent
current takes the form
\begin{equation}
    \langle j_y \rangle =\frac{f D_0^2}{D_0^2 \rho^2 (1-\rho/\rho_c)^2+f^2} \rho^{5/2} (1-\rho/\rho_c)^2.
    \label{eq:current_expression}
\end{equation}
The resonance in the particle current appears at the ratcheting rate 
$f=\nu=D_0 \rho (1-\rho/\rho_c)$ and the density dependent amplitude
take the form
$\kappa \rho v_0 = \kappa D_0 \rho^{3/2}(1-\rho/\rho_c)$. 

\begin{figure}[!t]
  \centering
  \includegraphics[width=0.7\linewidth]{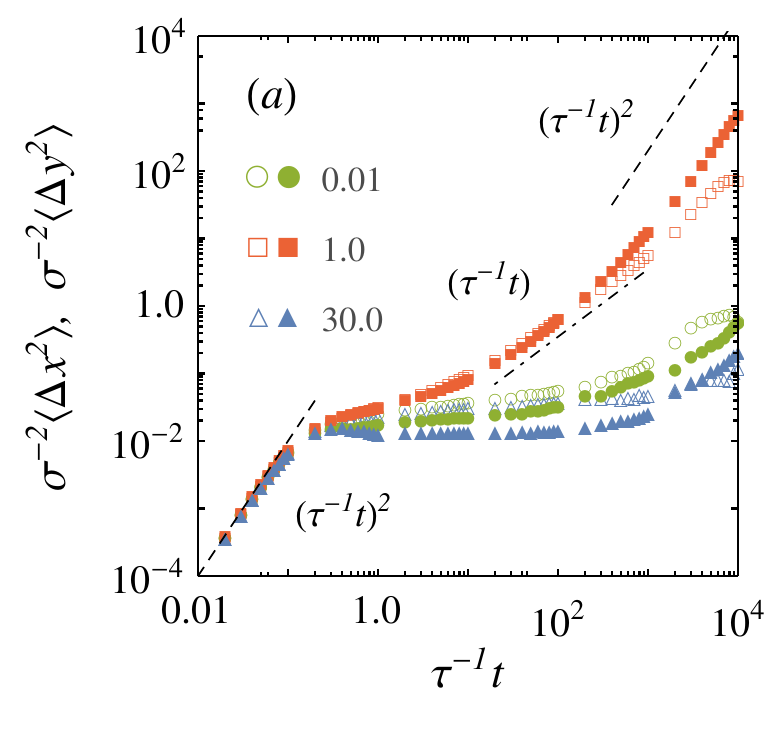}
    \includegraphics[width=\linewidth]{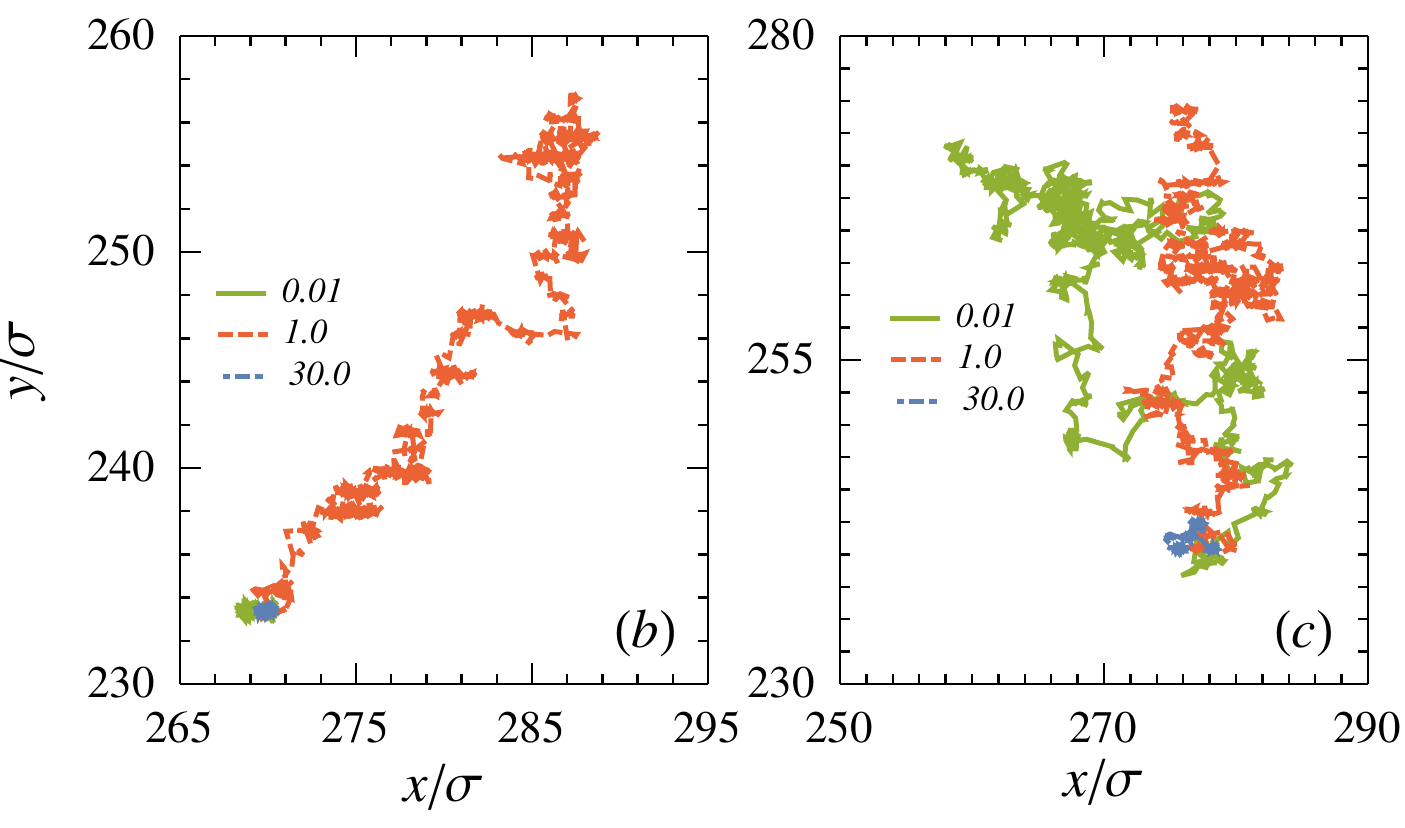}
    \caption{(color online
      ) 
      Plot of the mean-square displacement at density $ \rho \s^2 = 1.04$
      for frequencies $f \tau=0.01$ (green), $1.0$ (red) and $30.0$
      (blue).
    %   {\color{red} Use the same color-codes in ($a$), ($b$) and ($c$).} 
      The displacements in $\la \D x^2 \ra$ are shown with open
      symbols and that in $\la \D y^2 \ra$ are represented with filled
      symbols.  Typical trajectories of particles at densities
      $\rho\sigma^2=1.04$ ($b$) and $0.98$ ($c$) are shown for three
      different frequencies $f\tau=0.01$, $1$ and $30$. At high
      frequencies the particles get localized, whereas at very low
      frequencies localization happens only at densities above
      $\rho_c$. In the intermediate frequencies, at both densities the
      trajectories span a length scale of $\approx 20\,\s - 30\,\sigma$.
    }
  \label{fig:msdplot}
\end{figure}

\cref{fig:fluxplot}($a$) shows data collapse of particle currents when plotted 
as $(\kappa \rho v_0)^{-1} \langle j_y \rangle $ against the dimensionless variable $f/\nu$. The current maximizes at the resonance
frequency of $f=\nu$. 
\cref{fig:fluxplot}($b$) shows the limit of validity of the approximate form  $v_0 = D_0 \rho^{1/2}(1-\rho/\rho_c)$. 
A comparison of \cref{fig:phase_diagram} and
\cref{fig:fluxplot}$a$ shows the relationship between the structure
and dynamics. For example, at the resonance frequency, the system
melts in order to carry the largest directed current.

The dynamics at the local scale can be better appreciated by examining
trajectories of individual particles. First we consider the system at
high density $\rho \sigma^2=1.04$.  
Both the components of the  displacement fluctuations in \cref{fig:msdplot}($a$) show an initial ballistic part $\sim t^2$ due to the inertial nature of the dynamics.
Both in the low and high frequency limits, this crosses over to a sub-diffusive regime  ($\sim t^\nu$ with $\nu<1$). The test particles show  localized motion within cages formed by neighbours  (see \cref{fig:msdplot}($b$)\,).  
The intermediate frequency driving at $f\t=1$   shows a long time diffusive behavior in $x$, $\la \D x^2 \ra \sim t$, and almost ballistic motion in $y$, $\la \D y^2 \ra \sim t^\nu$  with $1 < \nu \lesssim 2$~(\cref{fig:msdplot}($a$)\,).  
The corresponding particle trajectories display long excursions in the $y$-direction, as is
displayed in \cref{fig:msdplot}($b$).  

\cref{fig:msdplot}($c$) shows typical particle trajectories at low densities, $\r \s^2 = 0.98$. They get localized only at the highest
driving frequencies $f \t=30$.  At small and
intermediate frequencies they show long excursions which are extended mainly in the $y$-direction, the direction of ratchet drive.  
The localization of trajectories is related to the low directed current carried by the system, and its ordering into a solid phase.  
Similarly, the extended particle trajectories are related to the melting of the solid and the presence of relatively large directed current.

\begin{figure}[!t]
    \centering
    \includegraphics[width=8.6cm]{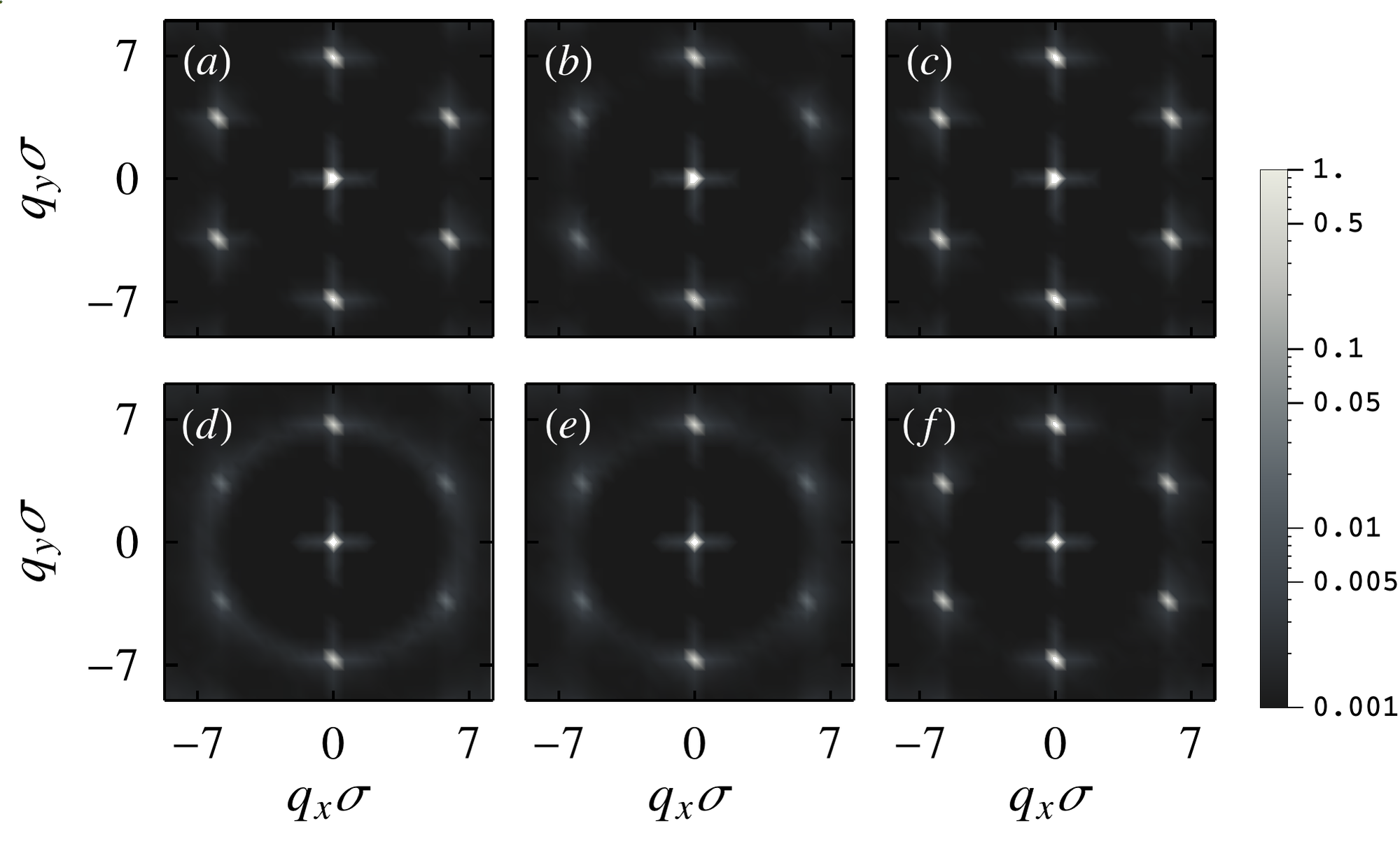}
    \caption{Plots of the static structure factor $\la \psi_{\bf q} \ra$ for the densities $\rho \sigma^2=1.04$ ($a$-$c$) and $0.98$ (figures $d$-$f$). 
    The three columns
      correspond to three different frequencies $f \tau=0.01$ ($a$ and $d$), $1.00$ ($b$ and $e$) and $30$ ($c$ and $f$).}
    \label{fig:struc}
\end{figure}

\subsection{Non-equilibrium melting}
\label{ssec:reentrant_transition}

For a more quantitative analysis, we turn our attention to the phase
diagram \cref{fig:phase_diagram}($a$) and consider the phase behavior
along constant frequency, and constant density lines.  The structure
factor,
$\la \psi_{\bf q} \ra = (1/N)\,\la \rho_{\bf q}\, \rho_{-\bf q} \ra$
(see \cref{fig:struc}) with
$\rho_{\bf q}=\sum_{j=1}^{N} e^{-\complexi {\bf q} \cdot \bf{r}_j}$
and $\rho^*_{\bf q}=\rho_{-\bf q}$, can clearly distinguish between a solid, hexatic, liquid and a modulated liquid phase~\cite{Chaikin2012}.  
In the solid phase $\la \psi_{\bf q} \ra$ shows a characteristic six fold symmetry with peaks at $\G_1=(0,\pm 2 \pi /a_y)$ and $\G_2=(\pm 2\pi/a,\pm \pi/a_y)$ reflecting the underlying triangular lattice structure~(see \cref{fig:struc}($a$),($c$),($f$)\,). 
The six intensity maxima broaden along the constant radius $q=2\pi/a$ circle in a hexatic~(\cref{fig:struc}($b$),($e$)\,). 
In a simple liquid with spherical symmetry, the broadening extends to overlap forming a characteristic ring structure. 
On the other hand, in a modulated liquid phase, $\la \psi_{\bf q} \ra$ is expected to show two bright spots at $\G_1$, in addition to the ring structure characterizing a simple liquid~(see \cref{fig:struc}($b$),($e$)\,). %
Note that the presence of the external periodic potential in the present context induces an  explicit symmetry breaking by imposing density modulations in $y$-direction, ensuring $\la \psi_{\G_1} \ra> \la \psi_{\G_2} \ra$.
The other four quasi Bragg peaks at $\G_2$, e.g., at the highest ratcheting frequencies, identify the appearance of the quasi long ranged positional order (QLRO). 
We use their arithmatic mean as the measure of solid order parameter $\la \psi_{\G_2} \ra$. 

\begin{figure}[!t]
  \centering
  \includegraphics[width=6cm]{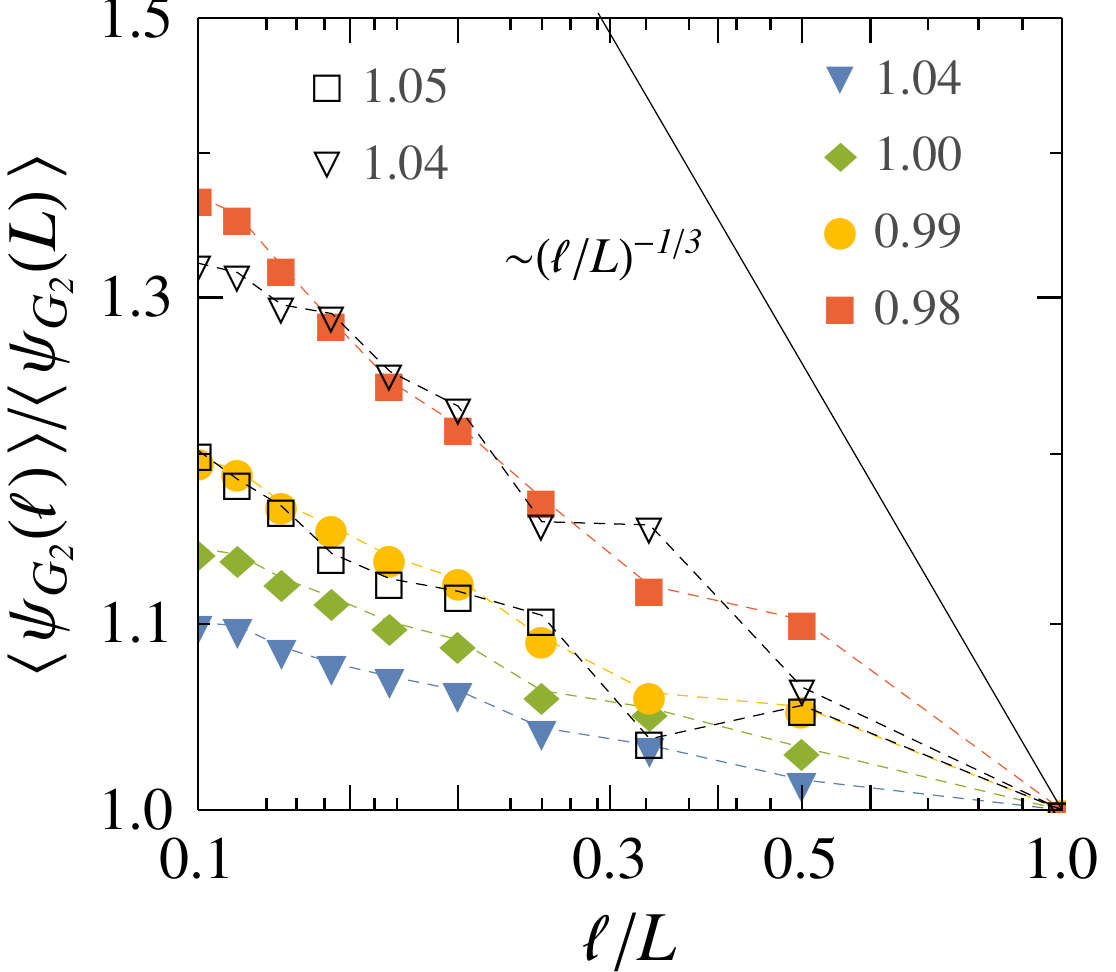}
  \caption{System size dependence of the solid order parameter. The ratio of mean order parameter measured over blocks of size $\ell= (\ell_x \times \ell_y)^{1/2}$ with respect to that measured over the whole system,
    $\la \psi_{\mathbf{G}_2}(\ell)\ra/\la \psi_{\mathbf{G}_2}(L) \ra$ decays with
 $\ell/L$ with power law $(\ell/L)^{-\nu}$.
 The data is shown for different densities as
indicated in the legend. The solid and open symbols denote results at frequency
    $f\t=30$ and $f\t=0.01$, respectively. 
    The solid black line is a plot of the power-law  $(\ell/L)^{-1/3}$ expected at the equilibrium KTHNY melting point.}
\label{fig:scaling_solid_order_parameter}
\end{figure}

The QLRO in the solid phase is explicitly demonstrated using the system size dependence of $\la \psi_{\G_2} \ra$ shown in
\cref{fig:scaling_solid_order_parameter}. The calculations are performed over sub-blocks of sizes  $\ell_x \times \ell_y = \ell^2$. 
Within both the high and low frequency solid regions, $\la \psi_{\G_2}(\ell) \ra \sim \ell^{-\nu}$ where $\nu < 1/3$, the value of the exponent expected at the equilibrium KTHNY (Kosterlitz- Thouless- Halperin-
Nelson- Young)
melting~\cite{Kosterlitz1973, Halperin1978, Young1979}. In this case, the value of the exponent $\nu$ depends on the mean density and ratcheting frequency.

\begin{figure}[!h]
  \centering
  \includegraphics[width=\linewidth]{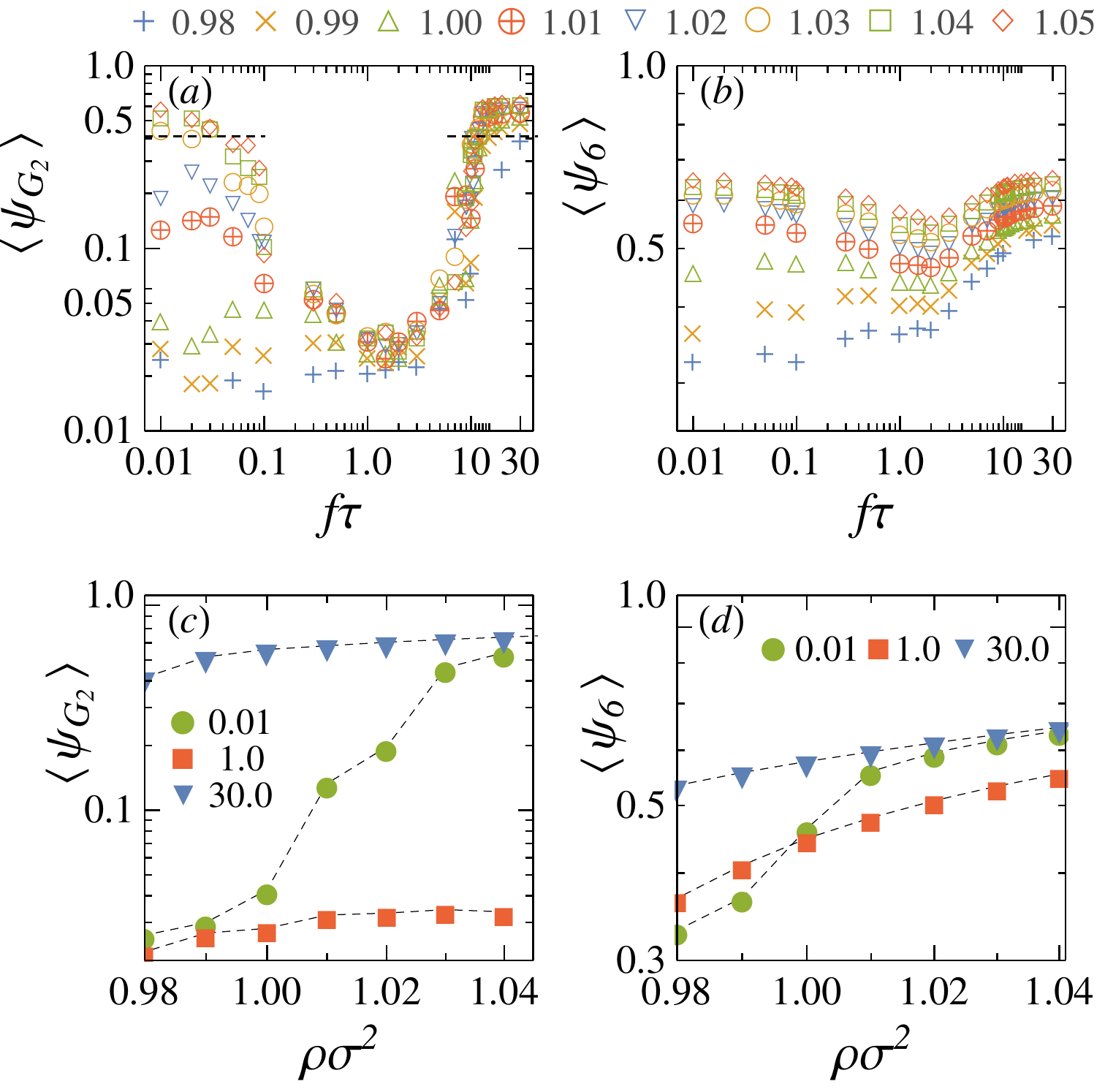}
\caption{ Dependence of the ($a$)\,solid order parameter $\langle \psi_{\bf{G}_2} \rangle $  and ($b$)\,hexatic order parameter $\langle \psi_6 \rangle$  as a function of frequency is  plotted for different densities as indicated in the legend on the top. 
The equilibrium melting point $\la \psi^{m}_{\bf{G}_2} \ra = 0.376$ is denoted by the dashed lines in ($a$). The variations of   ($c$)\,$\langle \psi_{\bf{G}_2} \rangle$ and  ($d$)\,$\langle \psi_6 \rangle$ are shown as a function of the mean density for three driving frequencies $f\tau = 0.01$, $1$ and $30$ as indicated in the legends. }
  \label{fig:solid_op_freq_density}
\end{figure}
 The phase behaviors are further characterized by following the change
in the hexatic bond orientational order
$\la \psi_6 \ra = (1/N) \la | \sum_{i=1}^N \ \vec \psi^i_6|^2 \ra$,
where we define the local hexatic order
$\vec \psi^i_6 = \sum_{k=1}^{n_v} (\ell_k/\ell) \exp(\complexi 6 \phi_{ik}
)$
utilizing the $n_v$ Voronoi neighbors of the $i$-th test particle. 
Here $\phi_{ik}$ is the angle subtended by the bond between the $i$-th particle and its $k$-th Voronoi neighbor.  
In this definition we used the weighted average over the weight factor $\ell_k/\ell$ such that
$\ell = \sum_{k=1}^{n_v} \ell_k$ and $\ell_k$ denotes the length of the Voronoi edge corresponding to the $k$-th topological neighbor~\cite{Mickel2013}.
  
In \cref{fig:solid_op_freq_density}($a$) and ($b$) we show the variations of
the solid and hexatic orders, $\langle \psi_{\G_2} \rangle$ and $\langle \psi_6 \rangle$, as a
function of the ratcheting frequency keeping the density of the system
fixed.  \cref{fig:solid_op_freq_density}($c$) and ($d$) show similar
plots, but as a function of the density, keeping the driving frequency fixed. 
Both the order parameters show non-monotonic variation with frequency. 
The large variation of the solid order parameter $\la \psi_{\G_2} \ra$ with $f\t$ signifies melting, followed by a {\em re-entrant
solidification}.  
Here we use the value of the solid order parameter at the equilibrium melting point, $\la \psi^m_{\G_2} \ra \approx 0.376$~(\cref{appendix:PT}), to identify the boundary of solid phase,  $\la \psi_{\G_2}\ra \ge \la \psi^m_{\G_2} \ra$. 
As we show later, the density-density correlation changes from a power law to an exponential decay across this melting.

As \cref{fig:solid_op_freq_density}($a$) shows, at all the densities
considered the system remains in a solid phase at the highest driving
frequencies. With reduction of frequency below $f\t \approx 10$, the
system melts. At densities $\r \s^2 \gtrsim 1.01$ the system
re-solidifies as the frequency is lowered further below
$f\t \approx 0.1$. At intermediate to high frequencies, whether the
system remains in a solid or fluid phase is essentially determined by
the driving frequency, and not the mean density of the system~(see
\cref{fig:solid_op_freq_density}($c$)). Only at the lowest driving
frequencies, one finds density dependent solidification.  
The hexatic order parameter $\la \psi_6\ra$ shows similar variations, albeit with lesser magnitude~(see \cref{fig:solid_op_freq_density}($b$), ($d$)\,). 
 At driving frequencies much larger than the inverse relaxation times, the system responds to only a time-integrated potential profile. The corresponding behavior is similar to that in the presence of a  time-independent commensurate potential. 
For equilibrium melting of the free system, and melting in the presence of a time-independent periodic potential commensurate with the density, see \cref{fig:op_free_lif} presented in
\cref{appendix:PT}.

 \begin{figure}[!t]
  \centering
 \includegraphics[width=\linewidth]{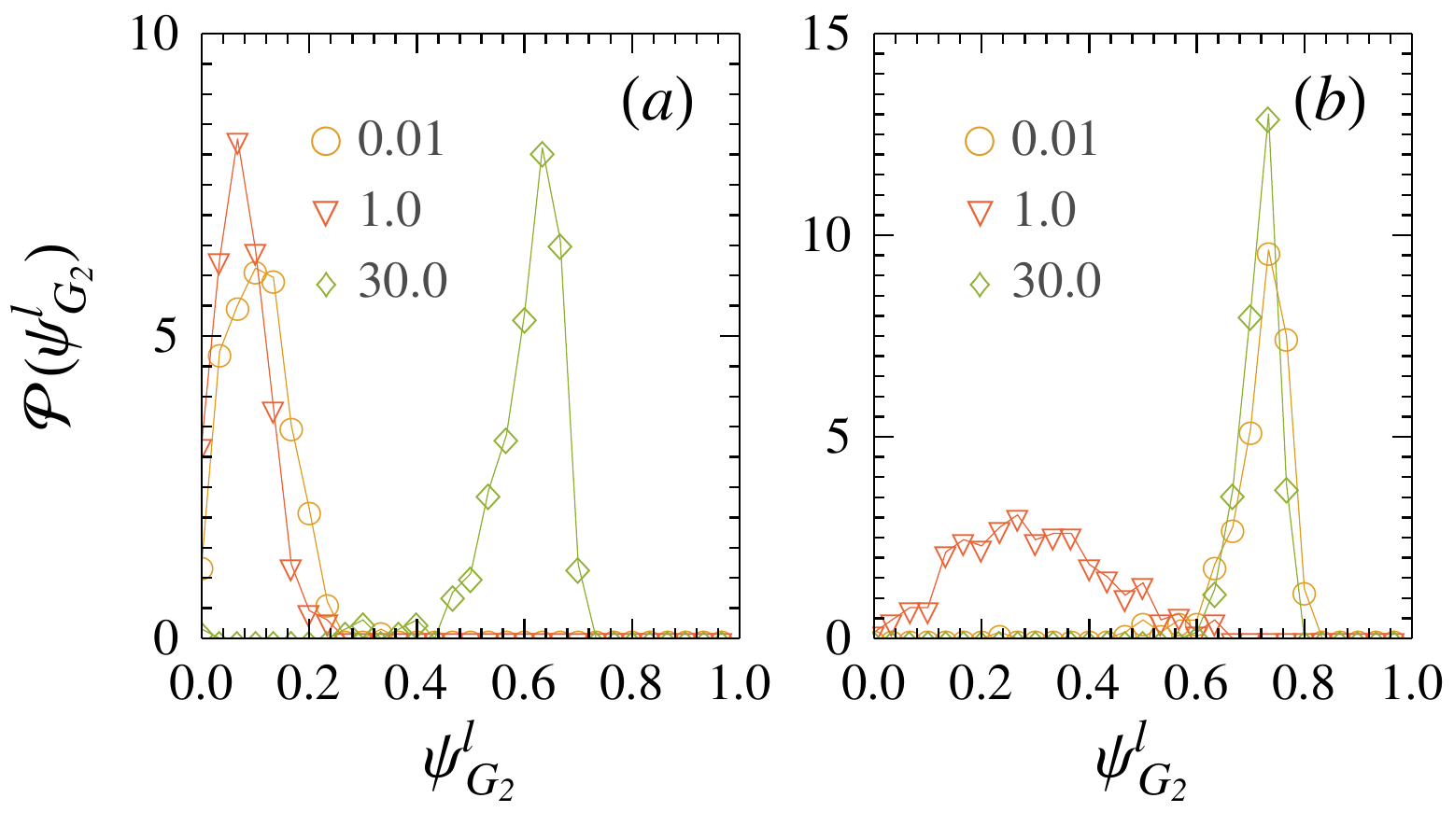}
 \includegraphics[width=\linewidth]{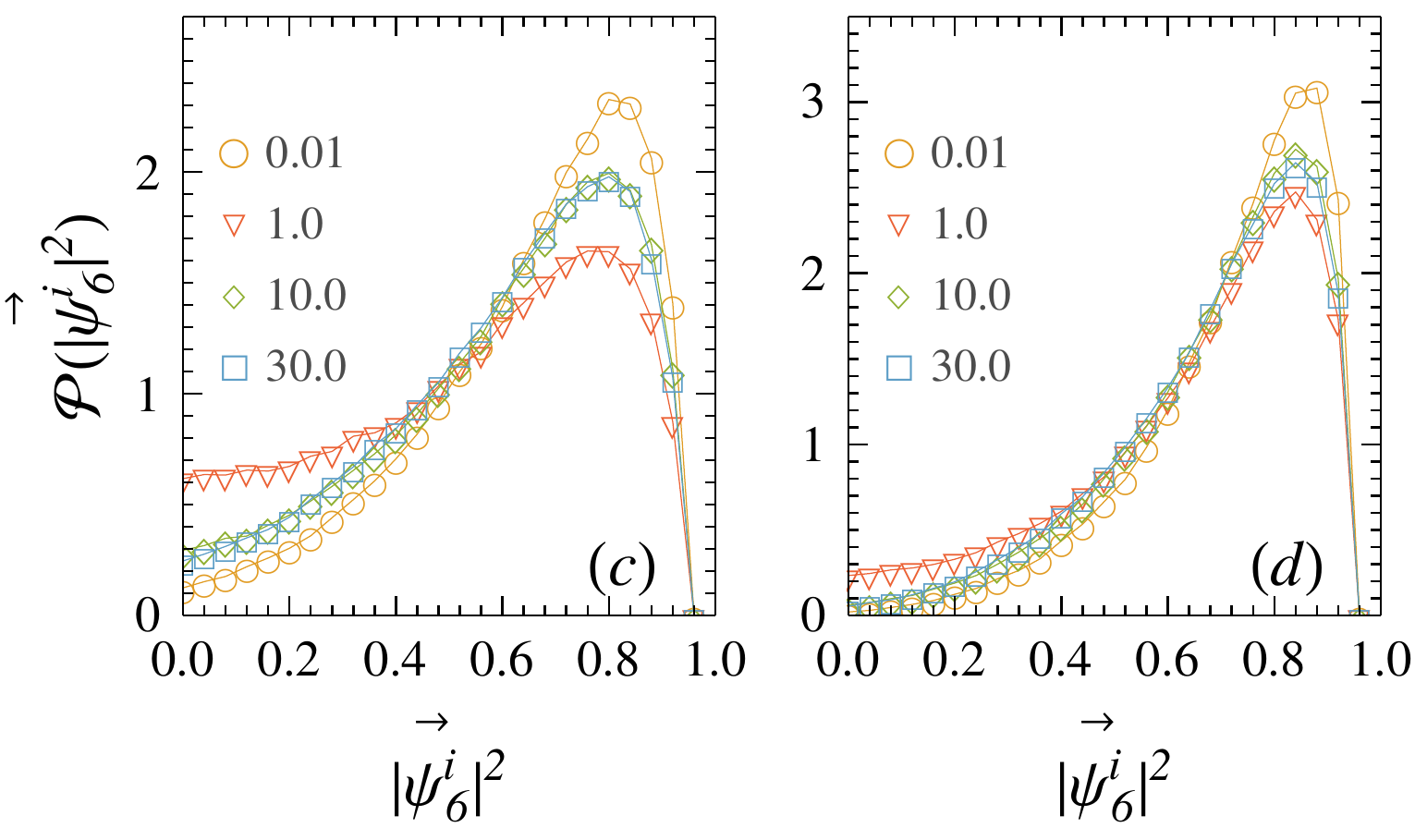}
 \caption{Probability distributions of the local solid and hexatic order parameters $\psi^l_{\bf{G}_2}$ and $|\vec \psi_6^i|^2$. The distribution functions ${\cal P}(\psi^l_{\bf{G}_2})$ at densities $\rho \sigma^2=0.98$\,($a$) and $\rho \sigma^2=1.04$\,($b$). The local solid order parameter is determined over subsystems of length $\ell = (\ell_x \times \ell_y)^{1/2}$, where $\ell_x/L_x=\ell_y/L_y=1/14$. 
 The distribution functions of local hexatic order ${\cal P}(|\vec \psi_6^i|^2)$ at $\rho \sigma^2=0.98$\,($c$) and $1.04$\,($d$), corresponding to four representative ratcheting frequencies as indicated in the
   legends.}
\label{fig:solid_op_hist_sq}
\end{figure}

\begin{figure}[!t]
  \centering
  \includegraphics[width=0.65\linewidth]{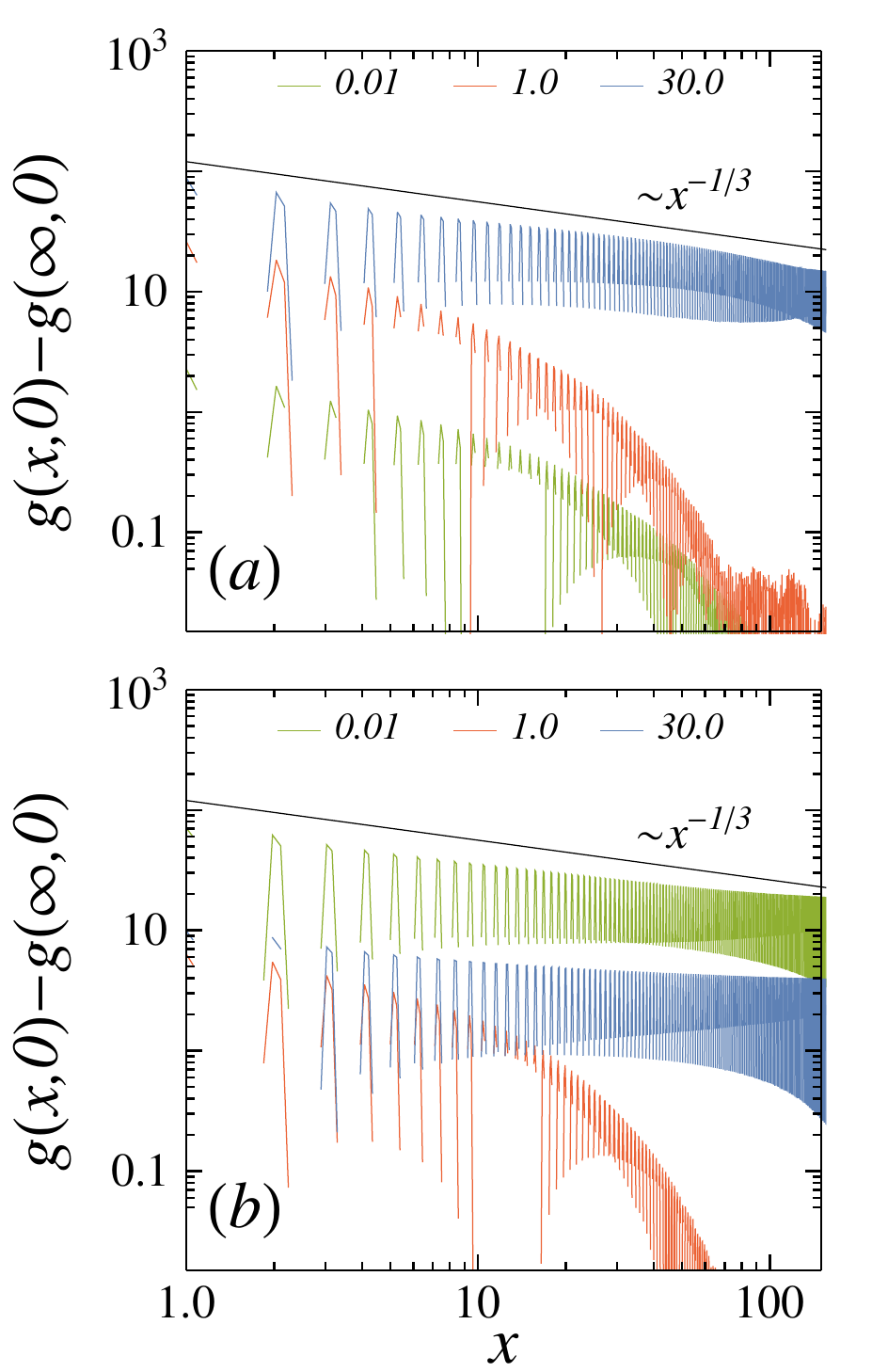}
  \caption{ (Color Online) 
  The pair correlation functions $g(x,0)-g(\infty,0)$ calculated at the mean 
    densities $\rho \sigma^2=0.98$ ($a$) and $1.04$ ($b$) for the frequencies indicated in the legends. 
    }
\label{fig:gx_plot}
\end{figure}

\subsection{Continuous transition: Distribution of order parameters}
%\dips{
We probe the order of the phase transitions using the distribution of
  the local solid and hexatic order parameters.  In determining the
  local solid order, we divided the simulation box into sub-boxes of
  size $\ell = (\ell_x \times \ell_y)^{1/2}$ with
  $\ell_x/L_x=\ell_y/L_y=1/14$.  $\psi^l_{\G_2}$ is then calculated
  using the definition of $\psi_{\bf q}$ restricted within these
  sub-boxes. For the local hexatic order parameter, we calculate 
 %\delete{ $|\rho_6(\bf{r}_i)|^2$} 
 $|\rho^i_6|^2$ for all the particles.  The distribution functions of these quantities, 
  ${\cal P}(\psi^l_{\G_2})$ and ${\cal P}(|\rho_{6}|^2)$ are plotted
  in \cref{fig:solid_op_hist_sq}.  They remain unimodal at all
  points of the phase diagram.  At low densities, $\r \s^2 = 0.98$ in
  \cref{fig:solid_op_hist_sq}($a$), the maximum of
  ${\cal P}(\psi^l_{\G_2})$ appears at an order parameter
  corresponding to the solid phase only at the highest frequencies.
  With decreasing frequency the peak shifts towards lower values,
  signifying melting below $f\t = 10$ and remain low at the lowest
  frequencies.  The unimodal nature of the distribution function, as
  the peak shifts to lower values, signifies the absence of any
  metastable state across the transition, a characteristic of
  continuous transitions.  At high densities, e.g., $\r \s^2 = 1.04$
  in \cref{fig:solid_op_hist_sq}($b$), corresponding to the
  re-entrant transition, the peak of the distribution
  ${\cal P}(\psi^l_{\G_2})$ shifts from a high to low and back to high
  values as the ratcheting rate decreases from the highest
  frequencies. As before, the unimodal nature of the distribution
  corresponds to a continuous melting transition.

\begin{figure*}[!t]
\centering
\includegraphics[width=0.95 \textwidth]{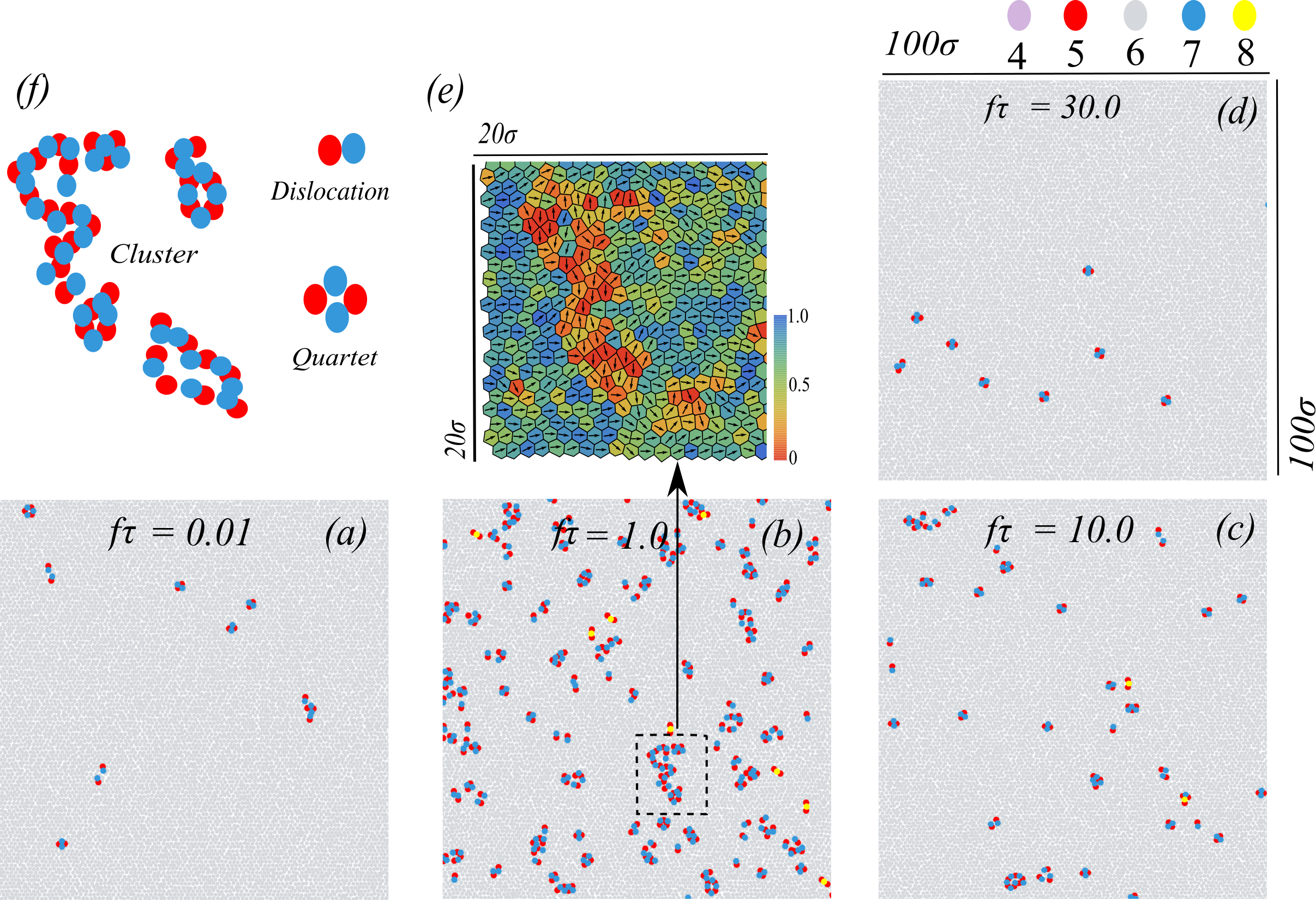}
\caption{
{($a$)-($d$)}: Configurations of the system in
  $100\sigma \times 100\sigma$ sub-volumes at a density of $\rho \sigma^{2}=1.04$ for frequencies $f \tau= 0.01$ ($a$), $1$ ($b$),
  $10$ ($c$) and $30$ ($d$). The particles are color coded according to the number of their topological neighbors $n_v$, with gray\,($n_v=6$), blue\,($n_v=7$), red\,($n_v=5$), purple\,($n_v=4$), and yellow\,($n_v=8$), as shown in the legend on top of ($d$).  
  {($e$)}\,The figure highlights the hexatic order in a region containing a connected string of defects marked by a dashed-box of size $20 \s \times 20 \s$ in ($b$). The Voronoi tesselations are shown for each particle within the box and are color coded according to the local value of the hexatic order. The arrows denote the orientation of the local hexatic order $\vec \psi_6^i$. {($f$)}\,Indicate typical examples of defect quartets, dislocations and defect clusters that appear in the system.}
\label{fig:defect_types}
\end{figure*}

The solid melts to a hexatic phase, characterized by the finite
hexatic order.  However, a further melting of the hexatic is not
observed as the frequency is varied. In the density-frequency range
bounded between the two dashed lines with open inverted triangles and
open circles denoted in \cref{fig:phase_diagram}~($a$), the system
remains in a hexatic phase. This is corroborated by the distribution
of the local hexatic order 
${\cal P}( |\rho_6|^2)$ shown in
\cref{fig:solid_op_hist_sq}($c$) and ($d$) corresponding to densities
$\rho \sigma^{2}=0.98$ and $1.04$, respectively.  The uni-modal nature
of the distribution with a roughly unchanged peak position and a fat
tail persists throughout the frequency range. The peak of the
distribution does not shift.  However, it is important to note that deep
inside the hexatic phase, near $f \t = 1$, a significant fraction of
the system displays vanishing hexatic order. This is more prominent at
lower densities~(see \cref{fig:solid_op_hist_sq}($c$)\,).  As we show
in a later section, local dip in the hexatic order is associated with
the formation of grain boundaries.
%}

\subsection{Melting of solid: Correlation functions} 
\label{ssec:pair_correlation}
The pair correlation functions $g(x,y)$ capture the solid melting~(see \cref{appendix:pair_correlation_function} for further details).   
A density modulation is externally induced in the system along the $y$-direction by the ratcheting drive, breaking the translational symmetry in that direction, explicitly. 
To study the spontaneous symmetry breaking, here we focus on the $x$-component of
the two point correlation functions $g(x,y)$. 
The component of the correlation 
$g(x,0)-g(\infty,0)$ along the axis
perpendicular to the direction of the ratcheting drive are shown in
\cref{fig:gx_plot}. This provides a more conclusive evidence to the
nature of the phases in different density and frequency regimes. 
At low densities, as we have shown before, the solid order exists only at the highest
frequencies. At such frequencies we find an algebraic decay of the
correlation along the $x$ direction, signifying the QLRO~(see
\cref{fig:gx_plot}($a$)\,).  At intermediate and low frequencies the
system melts. This is captured by the exponential decay of the
correlation with correlation length $\sim 10\sigma$~(see
\cref{fig:gx_plot}($a$)\,).  The scenario changes at higher
densities. At high frequency, as before, we again find a solid phase,
with the correlation exhibiting algebraic decay corresponding to the
QLRO~(\cref{fig:gx_plot}($b$)\,). At high densities, one obtains another 
solid phase at the low ratcheting frequencies. 
This also shows algebraic decay of correlations signifying QLRO~(see $f\t =0.01$ graph
in \cref{fig:gx_plot}($b$)\,).  
The power law $x^{-1/3}$ shown in the figures denote the expected correlation at the onset of  the equilibrium KTHNY melting.   
At the intermediate driving
frequencies, $f\t=1$ in $\r\s^2=1.04$ system, the correlation shows
exponential decay with a correlation length $\sim 10\,\s$, similar to the behavior observed in the low density regime~(\cref{fig:gx_plot}($b$)\,).
The change from algebraic to exponential decay is utilized to identify the solid melting points  in the phase diagram, as is detailed further in \cref{appendix:pair_correlation_function}.

\begin{figure*}[!t]
\centering
\includegraphics[width=0.9 \textwidth]{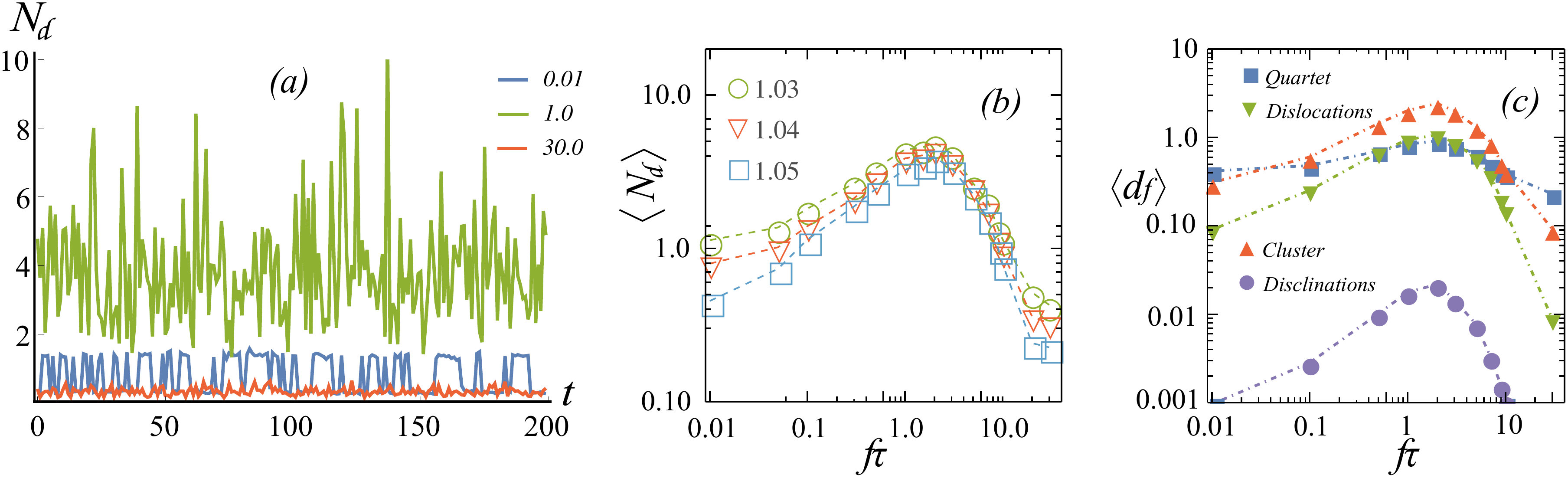}
\caption{{($a$)}\,Plot of the total defect percentage $N_d$ as a
  function of time for three different driving frequencies as
  indicated in the legend. The density of the system is
  $\rho \sigma^{2}=1.04$.  At high frequency the system remains in a solid phase, and the defect formation is significantly low ($N_d <1\%$). On the other hand, in the opposite limit
  of low frequency, $f \t=0.01$, the defect formation follows the swicthing of the driving potential, $N_d$ switches between two values. In the intermediate current carrying
  regime, the formation of the defects is maximized. The defects are
  further classified by types, e.g., quartets, dislocations, clusters, and disclinations.
 {($b$)}\,The average defect percentage $\la N_d \ra$ shows a non-monotonic variation with the ratcheting frequency. We show results at three densities denoted in the legend.
{($c$)}\,Variation of the time averaged percentage fraction of the individual defect types, the  quartets, dislocations, defect clusters, and disclinations with ratcheting frequency.}
\label{fig:defect_dynamics}
\end{figure*}

\subsection{Defect formation}
The equilibrium melting of the two- dimensional solid  within the
continuous KTHNY melting scenario is known to proceed by the unbinding
of dislocation pairs into free dislocations.  To identify such
topological defects, we first obtain the coordination number $n_v$ of
each particle in the system counting the number of its Voronoi
neighbours.  In a perfect triangular lattice $n_v=6$ for all
particles. We follow $n_v \neq 6$ particles to identify the $n_v$-fold
defects.  Even within the solid phase, fluctuations of bound quartets
of $5-7-5-7$ defects (bound dislocation pairs) keep appearing.  They
form dislocations by dissociating into separate $5-7$ and $7-5$
non-neighboring pairs.  Presence of a finite fraction of particles
associated with dislocations characterize the hexatic phase.
The system shows dislocation formation as the solid melts. Moreover,
we find defect clusters larger than quartets that are either compact
or string-like (grain boundary)~\cite{Qi:2014}. All the dominant
defect types observed in our simulations are indicated in
\cref{fig:defect_types}($f$). 
Their typical configurations in a sub-volume of size 
$100 \s \times 100 \s$ at $\r \s^2=1.04$ and different
ratcheting frequencies are shown in 
\cref{fig:defect_types}($a$)-($d$). In these figures,
the colors associated to particles indicate the number of
topological neighbors they have, $n_v=4\,$(purple), $5\,$(red), $6\,$(green), $7\,$(blue), $8\,$(yellow).
%with green color denoting particles with $n_v=6$ topological neighbors. 
Clearly, defect formation is suppressed at both the extremities of the
ratcheting frequency.  It increases significantly in the intermediate frequency
regime associated with solid melting (\cref{fig:defect_types}~($b$)).
The relative fraction of different defect types also vary with the
driving frequency. In the highest frequency solids, only bound
quartets (bound dislocation pairs) are observed in
\cref{fig:defect_types}~$(d)$. 
As the solid melts with decreasing frequency, dislocations and defect
clusters start to appear and eventually dominate over the quartets in
the system (\cref{fig:defect_types}~$(b)$ and ($c$)\,).  The
string-like defects remain extended along the $y$-direction, the
direction of particle current under ratcheting. Such a connected
string of defects is shown in \cref{fig:defect_types}~$(b)$ and has
been highlighted in \cref{fig:defect_types}~$(e)$, which shows Voronoi
diagram of a region containing the connected string of defects. The
color code in each Voronoi cell denotes the amount of hexatic order,
and the arrows denote the corresponding hexatic orientations.  At the location of the 
connected clusters of defects, the local hexatic order is low, and shows 
a hexatic orientation approximately orthogonal to the neighboring
defect-free regions.  With further lowering of the ratcheting frequency
below $f_s$, the defect fraction decreases strongly.

This description is quantified by focusing on the time evolution
of defect fractions.  We first consider the evolution of the total
fraction of all the topological defects, the percentage of particles
having non-six Voronoi neighbors $N_d = (1 - n_6/N)\times 100$, where
$n_6$ denotes the total number of particles with
$n_v=6$~(\cref{fig:defect_dynamics}~$(a)$\,).  Clearly, the largest
value of $N_d$ with the strongest fluctuations appear at the
intermediate frequencies. The defect formation gets dramatically
suppressed in the solid phase corresponding to the high ratcheting
frequencies.  At the lowest frequencies ($f\t=0.01$ in
\cref{fig:defect_dynamics}~$(a)$\,), $N_d$ remains relatively low and
follows the switching of external potential. 

The mean value $\la N_d \ra$ remains less than $4\%$ and varies
non-monotonically with $f\t$~(\cref{fig:defect_dynamics}~$(b)$\,).  It
shows a maximum at the resonance frequency corresponding to the
largest directed current, relating formation of topological defects
with carrying capacity of particle current in the system.

Further insight into the structure- dynamics relations can be obtained
by following the behavior of different defect fractions separately.
For this purpose, the percentage fraction $d_f$ of a defect type is
defined as $d_f=(n_d/N) \times 100$, where $n_d$ is the total number
of $n_v\neq 6$ particles that may contribute to either a quartet, a
dislocation, a cluster, or a disclination as described above.  The
time averaged percentage fractions of these topological defects as a
function of the driving frequency is shown in
\cref{fig:defect_dynamics}~$(c)$.  They exhibit a similar
non-monotonic behavior as $N_d$ and the mean particle current.  In the
high frequency solid, the dominant defects are the quartets.  As the
frequency is decreased, the melting of the solid is mediated by the
unbinding of these quartets into dislocations. The dislocation fraction becomes larger than that of quartets.  More importantly, at the resonance melting, the formation of defect
clusters dominate~(\cref{fig:defect_dynamics}$(c)$\,).   
The fraction of disclinations remain relatively
insignificant (less than $0.03\%$), about two orders of magnitude smaller than that of the defect clusters. This is consistent with the fact that the hexatic does not melt within these parameter regimes.

\section{Discussion}
\label{sec:outlook}
In conclusion, using a large scale simulation involving $262144$
particles, we have presented a detailed study of a ratcheted
two-dimensional colloidal suspension, focusing on the structure-
dynamics relationship.  The mean directed particle current driven by
the ratchet exhibits a resonance behavior.  Associated with this, the
solid melts to hexatic providing a mechanism allowing directed
transport.  The system exhibits a rich non-equilibrium phase diagram
as a function of the driving frequency and mean density. At high
densities, we found a re-entrant melting  transition as a function of ratcheting frequency.  The different phases are
characterized by the spatially resolved density profile, the
density-density correlation function, the structure factor, the solid and hexatic order parameters, and their distribution functions.
The role of the defects in the phase transition has been investigated in
detail.
The solid- melting is associated with formation of dislocations, but
unlike the equilibrium two- dimensional melting, this non-equilibrium melting is
dominated by the formation of defect clusters,  connected strings
of defects that remain oriented largely along the direction of the ratcheting 
drive.   
Remarkably, the driven hexatic does not melt to a fluid within the studied range of density and ratcheting drive.
Our detailed predictions regarding the variation of particle current and associated phase transitions can be verified using colloidal particles and optical~\cite{Faucheux1995} or magnetic ratcheting~\cite{Tierno2012} in a suitable laser trapping setup~\cite{Wei1998}. The impact of changing degree of potential asymmetry on the dynamics and phase behavior remains an interesting future direction of  study. 

\section*{Acknowledgements}
DC thanks ICTS-TIFR, Bangalore for an associateship, and SERB, India, for financial support through grant number EMR/2016/001454.

\appendix

\section{Relaxation under external potential}
\label{appendix:relaxtion}

\begin{figure}[!t]
    \centering
    \includegraphics[width=0.5\textwidth]{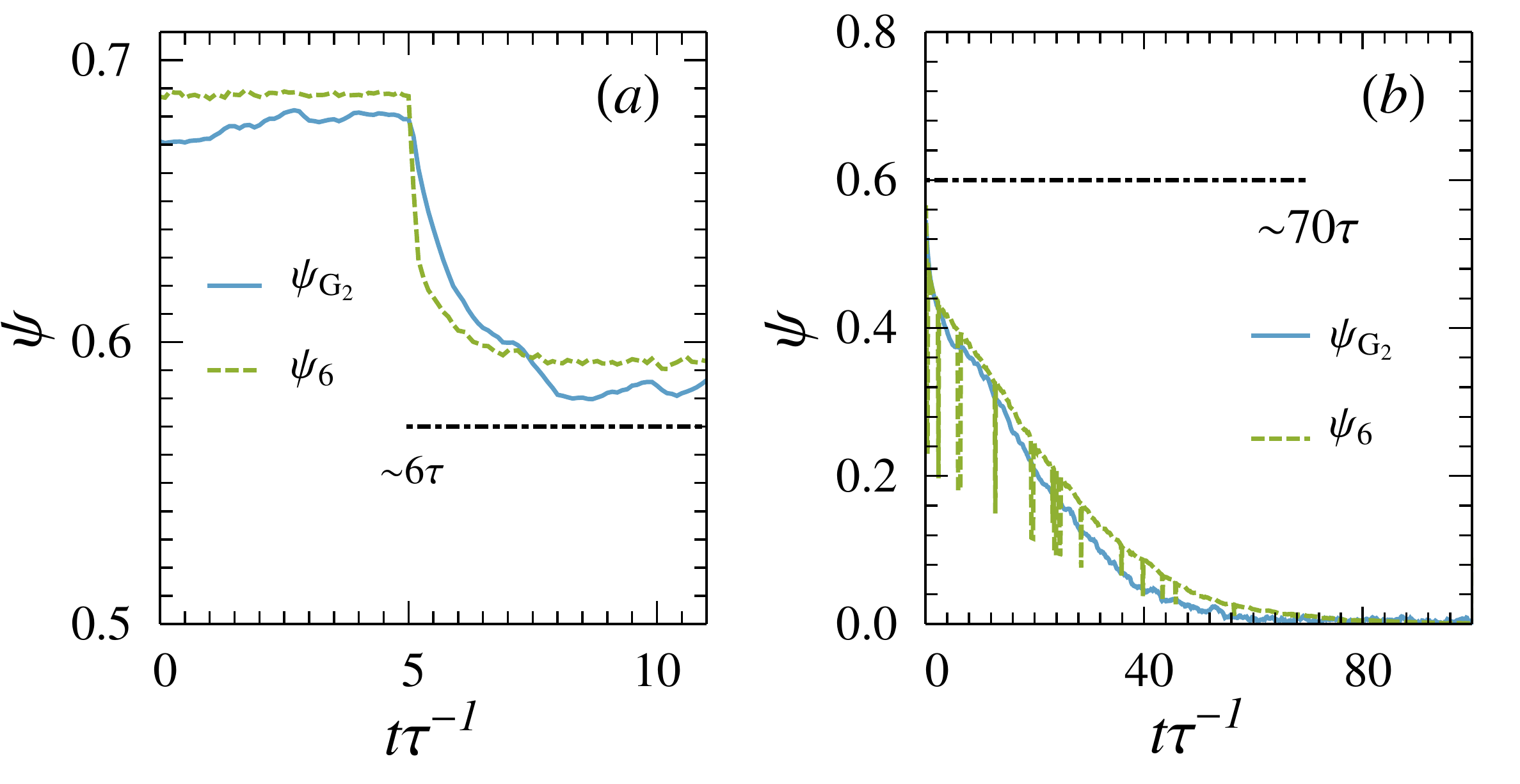}
    \caption{Relaxation dynamics of the solid and hexatic order parameters  at  $\r \s^2 =1.04$\,($a$) and $\r \s^2 =0.98$\,($b$), respectively. The time scales for relaxation are $f_s^{-1} \approx 6\,\t$\,($a$) and $f_s^{-1}\approx 70\,\t$\,($b$).}
    \label{fig:relaxation_frequency}
\end{figure}

\begin{figure}[!t]
\centering
\includegraphics[width=0.5\textwidth]{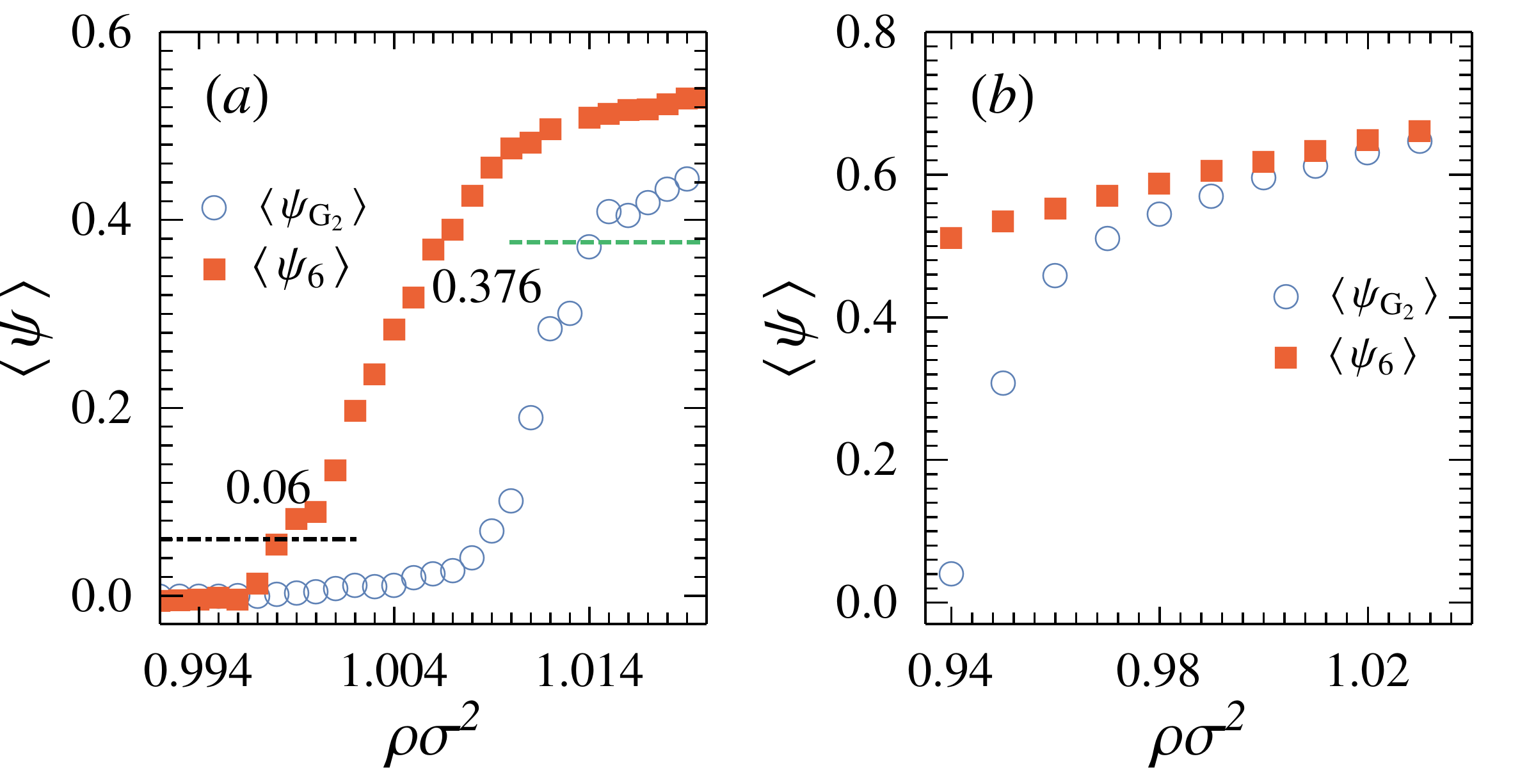}
\caption{Plot of the solid order parameter $\la \psi_{\bf{G}_2}\ra$ and the hexatic order parameter $\la \psi_6 \ra$ for the equilibrium phase transition and the laser induced freezing transition. The green dashed line denotes solid melting point $\la \psi_{{\bf G}_2}^m \ra=0.376$, and the black dash-dotted line denotes the hexatic melting point $\la \psi_6 \ra=0.06$.}
 \label{fig:op_free_lif}
\end{figure}
%%%

In this section, we show the results for the relaxation time-scales of
the solid and the hexatic order parameters after withdrawing an external potential commensurate with the system density under which the system is initially equilibrated. 
This time-scale at different densities are determined from separate simulations. 
The initial equilibration is performed under a  time-independent external potential of the form given in \cref{sec:model_simulation} with $V(t)=\epsilon$ over $10^7$ simulation steps.  
Thereafter, the external potential is removed and the evolution of the solid and the hexatic order parameters are measured over time.  

In \cref{fig:relaxation_frequency} we show the time evolution of these quantities at mean densities  $\rho \sigma^2=1.04$ (figure $a$) and $0.98$ (figure $b$). 
At densities higher than the equilibrium melting point $\r_m \s^2 \approx 1.01$, the solid and the hexatic order parameters, $ \psi_{\G_2}$ and $\psi_6$, decay to  finite values, indicating order even in the absence of external potential~(\cref{fig:relaxation_frequency}($a$)\,). 
In contrast, at densities $\r < \r_m$, the solid and the hexatic order parameters vanish with time~(\cref{fig:relaxation_frequency}($b$)\,). %
The time-scale of such decay, indicated in \cref{fig:relaxation_frequency}, gives the  estimate of the relevant relaxation time. 
If a time-independent external potential switches with a rate slower than this time-scale, the system will have enough time to {\em equilibrate} to instantaneous potential profiles.    
The relaxation frequency $f_s$ is inverse of  this time-scale, and has been indicated by the dash-dotted line through $\blacksquare$ symbols in the phase diagram \cref{fig:phase_diagram}.

\section{Equilibrium Phase Transition}
\label{appendix:PT}
One way we identified the solid melting in the main text is by
choosing an appropriate cutoff value of the solid order parameter $\la \psi_{\G_2} \ra$.  
For this, we used the equilibrium melting point in the absence of external potential.  
To demonstrate this, we perform separate molecular dynamics simulations of particles interacting via the soft-core 
 potential.
The temperature of the system is kept fixed at $T=1.0 \e/\kb$ using a Langevin heat bath.
The corresponding variation of $\la \psi_{\G_2}\ra$ and $\la \psi_6 \ra$ with the mean density is shown in \cref{fig:op_free_lif}($a$).  
The order parameters change continuously from low to high values. 
The solid melting point is separately determined by following the pressure-density curve and the density correlation function as in Ref.\citenum{Kapfer:2015ca} (data not shown).
This melting point is found at density $\rho_m \sigma^2 \approx 1.014$, where the solid order parameter $\la \psi_{\G_2}^m\ra \approx 0.376$.  
The hexatic order remains significantly large even at lower densities.  
The melting of hexatic to liquid is identified from the change in  correlation of the hexatic order parameter $\la g_6(r) \ra$, which transforms from an algebraic decay in the hexatic phase to an exponential decay in the liquid.  
The hexatic- melting is obtained at $\rho \sigma^2 \approx 0.998$, where $\la \psi_6 \ra \approx 0.06$. 
In \cref{fig:op_free_lif}($b$) we show a similar plot for the two order parameters but in the presence of a time- independent potential of the form $U_{ext}(x,y)$ given in \cref{sec:model_simulation} with $V(t)=\epsilon$.
In the presence of this potential,  the hexatic does not melt in the regime of $\r \s^2 \geq 0.94$.
The external potential maintains a significant hexatic order, with a value greater than $\la \psi_6 \ra =0.51$, although the solid order does drop below $0.376$ at $\r \s^2 \approx 0.96$.

\begin{figure}[!t]
\centering
\includegraphics[width=0.95\linewidth]{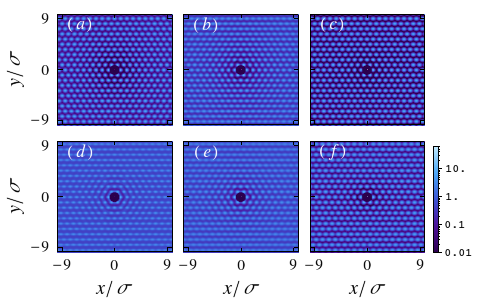}
\caption{  Pair correlation functions in the driven 2D colloidal suspension at
   densities $\rho\sigma^2=1.04$ in the top panels ($a$-$c$) and $\rho\sigma^2$=0.98 in the bottom panels ($d$-$f$) corresponding to three different frequencies $f\tau=0.01$ (left
  column), $f\tau=1$ (middle column) and $f\tau=30$ (right
  column). 
  }
\label{fig:gxy_plot}
\end{figure}
\section{Non-equilibrium melting and pair correlation}
\label{appendix:pair_correlation_function}
The two dimensional pair correlation functions $g(x,y) = \la \r(x,y) \r(0,0)\ra/ \la \r \ra^2$ at $\rho \sigma^2=0.98$ and $\rho \sigma^2=1.04$ are shown in \cref{fig:gxy_plot} at the three representative frequencies: low ($f\tau=0.01$), intermediate ($f\tau=1$) and high frequency ($f\tau=30$). 
The figures show the correlations over a length scale of $\pm 9\s$. 
While the clear contrast in \cref{fig:gxy_plot}($a$), ($c$)
and ($f$) demonstrate the  triangular lattice symmetry, the local diffused approximately triangular structures in the other $g(x,y)$ figures are characteristic of the hexatic phase.

\begin{figure*}[!t]
\includegraphics[width=\textwidth]{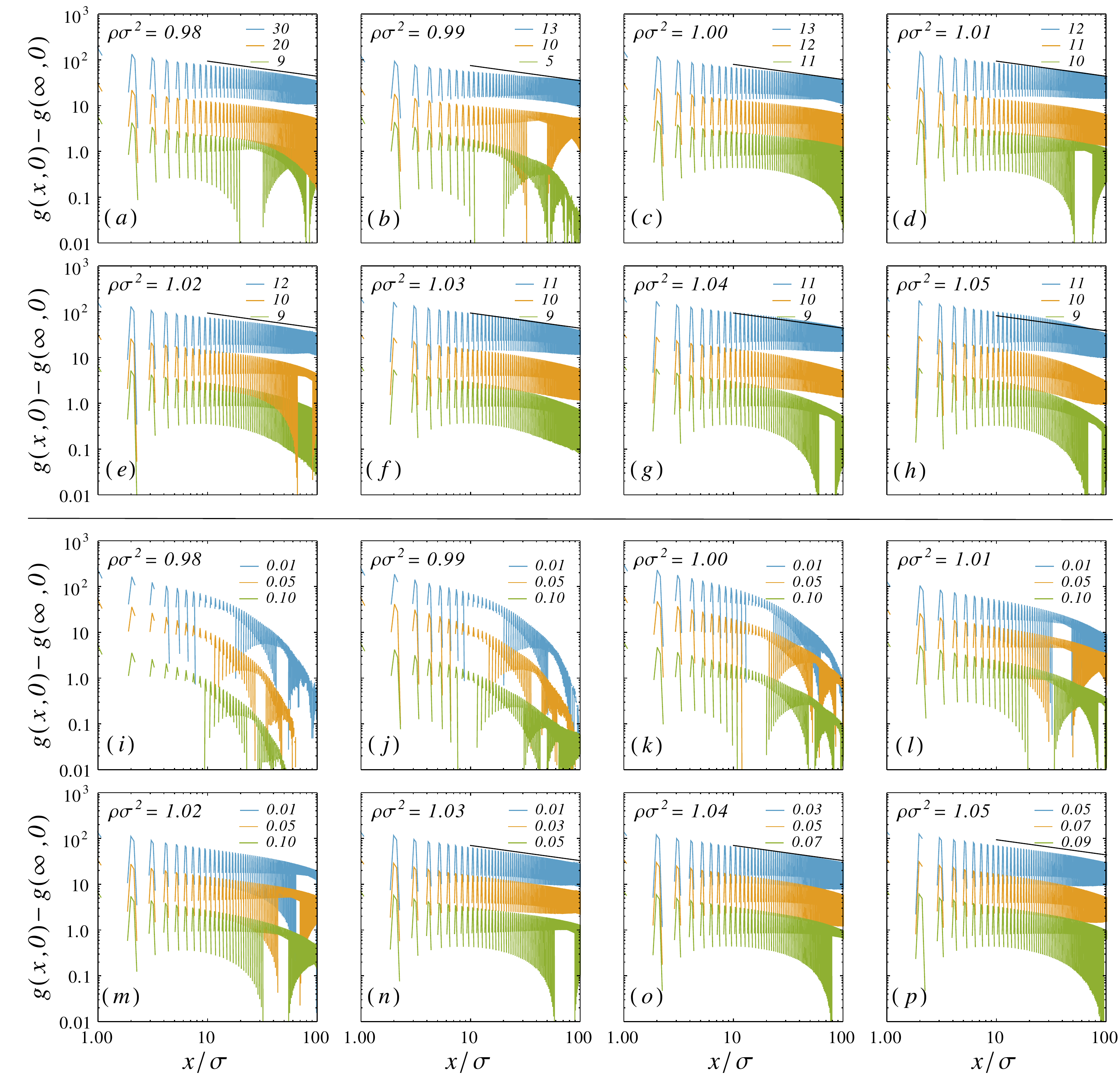}
\caption{Plots of $g(x,0)-g(\infty,0)$ at different densities  $\r \s^2$ of the system for frequencies $f \t$ indicated in the legends. In ($a$)--($h$)  we show the data near the high-frequency phase boundary, and in ($i$)--($p$) near the low-frequency phase boundary. 
The solid black lines denote the power law $x^{-1/3}$ that indicates the expected correlation at the solid melting point within KTHNY thoery~\cite{Kosterlitz1973, Halperin1978, Young1979}.}
\label{fig:gx_trans}
\end{figure*}

The component of the pair correlation function along the minima of the potential $g(x,0)-g(\infty,0)$ changes from a power-law to exponential decay with changing ratcheting frequency, identifying the melting point of a quasi- long ranged ordered solid. 
In \cref{fig:gx_trans}, $(a)$--$(h)$ show the high-frequency melting, while $(i)$--$(p)$ show possible melting at low frequencies.

At $\r \s^2 \lesssim 1.03$ and low frequencies, the decay of the pair correlation function remains always exponential, identifying an absence of transition.  
A crossover to an algebraic decay in $g(x,0)-g(x,\infty)$ appears at $\r \s^2 \gtrsim 1.03$, resulting in solid- hexatic transition points.
The phase boundaries displayed in \cref{fig:phase_diagram}~($a$) are consistent with the transition points obtained from this analysis.

%%%REFERENCES%%%

\providecommand*{\mcitethebibliography}{\thebibliography}
\csname @ifundefined\endcsname{endmcitethebibliography}
{\let\endmcitethebibliography\endthebibliography}{}


\begin{mcitethebibliography}{39}
\providecommand*{\natexlab}[1]{#1}
\providecommand*{\mciteSetBstSublistMode}[1]{}
\providecommand*{\mciteSetBstMaxWidthForm}[2]{}
\providecommand*{\mciteBstWouldAddEndPuncttrue}
  {\def\EndOfBibitem{\unskip.}}
\providecommand*{\mciteBstWouldAddEndPunctfalse}
  {\let\EndOfBibitem\relax}
\providecommand*{\mciteSetBstMidEndSepPunct}[3]{}
\providecommand*{\mciteSetBstSublistLabelBeginEnd}[3]{}
\providecommand*{\EndOfBibitem}{}
\mciteSetBstSublistMode{f}
\mciteSetBstMaxWidthForm{subitem}
{(\emph{\alph{mcitesubitemcount}})}
\mciteSetBstSublistLabelBeginEnd{\mcitemaxwidthsubitemform\space}
{\relax}{\relax}

\bibitem[Julicher \emph{et~al.}(1997)Julicher, Ajdari, and
  Prost]{Julicher1997a}
F.~Julicher, A.~Ajdari and J.~Prost, \emph{Reviews of Modern Physics}, 1997,
  \textbf{69}, 1269--1282\relax
\mciteBstWouldAddEndPuncttrue
\mciteSetBstMidEndSepPunct{\mcitedefaultmidpunct}
{\mcitedefaultendpunct}{\mcitedefaultseppunct}\relax
\EndOfBibitem
\bibitem[Astumian and H\"{a}nggi(2002)]{Astumian2002}
R.~D. Astumian and P.~H\"{a}nggi, \emph{Physics Today}, 2002, \textbf{55},
  33\relax
\mciteBstWouldAddEndPuncttrue
\mciteSetBstMidEndSepPunct{\mcitedefaultmidpunct}
{\mcitedefaultendpunct}{\mcitedefaultseppunct}\relax
\EndOfBibitem
\bibitem[H\"{a}nggi(2009)]{Hanggi2009}
P.~H\"{a}nggi, \emph{Reviews of Modern Physics}, 2009, \textbf{81},
  387--442\relax
\mciteBstWouldAddEndPuncttrue
\mciteSetBstMidEndSepPunct{\mcitedefaultmidpunct}
{\mcitedefaultendpunct}{\mcitedefaultseppunct}\relax
\EndOfBibitem
\bibitem[Reimann(2002)]{Reimann2002}
P.~Reimann, \emph{Physics Reports}, 2002, \textbf{361}, 57--265\relax
\mciteBstWouldAddEndPuncttrue
\mciteSetBstMidEndSepPunct{\mcitedefaultmidpunct}
{\mcitedefaultendpunct}{\mcitedefaultseppunct}\relax
\EndOfBibitem
\bibitem[Brouwer(1998)]{Brouwer1998}
P.~Brouwer, \emph{Phys. Rev. B}, 1998, \textbf{58}, R10135--R10138\relax
\mciteBstWouldAddEndPuncttrue
\mciteSetBstMidEndSepPunct{\mcitedefaultmidpunct}
{\mcitedefaultendpunct}{\mcitedefaultseppunct}\relax
\EndOfBibitem
\bibitem[Citro \emph{et~al.}(2003)Citro, Andrei, and Niu]{Citro2003}
R.~Citro, N.~Andrei and Q.~Niu, \emph{Phys. Rev. B}, 2003, \textbf{68},
  165312\relax
\mciteBstWouldAddEndPuncttrue
\mciteSetBstMidEndSepPunct{\mcitedefaultmidpunct}
{\mcitedefaultendpunct}{\mcitedefaultseppunct}\relax
\EndOfBibitem
\bibitem[Jain \emph{et~al.}(2007)Jain, Marathe, Chaudhuri, and Dhar]{Jain2007}
K.~Jain, R.~Marathe, A.~Chaudhuri and A.~Dhar, \emph{Phys. Rev. Lett.}, 2007,
  \textbf{99}, 190601\relax
\mciteBstWouldAddEndPuncttrue
\mciteSetBstMidEndSepPunct{\mcitedefaultmidpunct}
{\mcitedefaultendpunct}{\mcitedefaultseppunct}\relax
\EndOfBibitem
\bibitem[Chaudhuri and Dhar(2011)]{Chaudhuri2011}
D.~Chaudhuri and A.~Dhar, \emph{EPL (Europhysics Letters)}, 2011, \textbf{94},
  30006\relax
\mciteBstWouldAddEndPuncttrue
\mciteSetBstMidEndSepPunct{\mcitedefaultmidpunct}
{\mcitedefaultendpunct}{\mcitedefaultseppunct}\relax
\EndOfBibitem
\bibitem[Chaudhuri \emph{et~al.}(2015)Chaudhuri, Raju, and Dhar]{Chaudhuri2015}
D.~Chaudhuri, A.~Raju and A.~Dhar, \emph{Phys. Rev. E}, 2015, \textbf{91},
  050103\relax
\mciteBstWouldAddEndPuncttrue
\mciteSetBstMidEndSepPunct{\mcitedefaultmidpunct}
{\mcitedefaultendpunct}{\mcitedefaultseppunct}\relax
\EndOfBibitem
\bibitem[Chaudhuri(2015)]{Chaudhuri2015f}
D.~Chaudhuri, \emph{J. Phys. Conf. Ser.}, 2015, \textbf{638}, 012011\relax
\mciteBstWouldAddEndPuncttrue
\mciteSetBstMidEndSepPunct{\mcitedefaultmidpunct}
{\mcitedefaultendpunct}{\mcitedefaultseppunct}\relax
\EndOfBibitem
\bibitem[Gadsby \emph{et~al.}(2009)Gadsby, Takeuchi, Artigas, and
  Reyes]{Gadsby2009}
D.~C. Gadsby, A.~Takeuchi, P.~Artigas and N.~Reyes, \emph{Philos. Trans. R.
  Soc. B Biol. Sci.}, 2009, \textbf{364}, 229--238\relax
\mciteBstWouldAddEndPuncttrue
\mciteSetBstMidEndSepPunct{\mcitedefaultmidpunct}
{\mcitedefaultendpunct}{\mcitedefaultseppunct}\relax
\EndOfBibitem
\bibitem[Faucheux \emph{et~al.}(1995)Faucheux, Bourdieu, Kaplan, and
  Libchaber]{Faucheux1995}
L.~Faucheux, L.~Bourdieu, P.~Kaplan and A.~Libchaber, \emph{Physical Review
  Letters}, 1995, \textbf{74}, 1504--1507\relax
\mciteBstWouldAddEndPuncttrue
\mciteSetBstMidEndSepPunct{\mcitedefaultmidpunct}
{\mcitedefaultendpunct}{\mcitedefaultseppunct}\relax
\EndOfBibitem
\bibitem[Lopez \emph{et~al.}(2008)Lopez, Kuwada, Craig, Long, and
  Linke]{Lopez2008}
B.~Lopez, N.~Kuwada, E.~Craig, B.~Long and H.~Linke, \emph{Physical Review
  Letters}, 2008, \textbf{101}, 220601\relax
\mciteBstWouldAddEndPuncttrue
\mciteSetBstMidEndSepPunct{\mcitedefaultmidpunct}
{\mcitedefaultendpunct}{\mcitedefaultseppunct}\relax
\EndOfBibitem
\bibitem[Tierno \emph{et~al.}(2010)Tierno, Reimann, Johansen, and
  Sagu\'{e}s]{Tierno2010}
P.~Tierno, P.~Reimann, T.~H. Johansen and F.~Sagu\'{e}s, \emph{Physical Review
  Letters}, 2010, \textbf{105}, 230602\relax
\mciteBstWouldAddEndPuncttrue
\mciteSetBstMidEndSepPunct{\mcitedefaultmidpunct}
{\mcitedefaultendpunct}{\mcitedefaultseppunct}\relax
\EndOfBibitem
\bibitem[Tierno(2012)]{Tierno2012}
P.~Tierno, \emph{Physical Review Letters}, 2012, \textbf{109}, 198304\relax
\mciteBstWouldAddEndPuncttrue
\mciteSetBstMidEndSepPunct{\mcitedefaultmidpunct}
{\mcitedefaultendpunct}{\mcitedefaultseppunct}\relax
\EndOfBibitem
\bibitem[Rousselet \emph{et~al.}(1994)Rousselet, Salome, Ajdari, and
  Prost]{Rousselet1994}
J.~Rousselet, L.~Salome, A.~Ajdari and J.~Prost, \emph{Nature}, 1994,
  \textbf{370}, 446\relax
\mciteBstWouldAddEndPuncttrue
\mciteSetBstMidEndSepPunct{\mcitedefaultmidpunct}
{\mcitedefaultendpunct}{\mcitedefaultseppunct}\relax
\EndOfBibitem
\bibitem[Leibler(1994)]{Leibler1994}
S.~Leibler, \emph{Nature}, 1994, \textbf{370}, 412\relax
\mciteBstWouldAddEndPuncttrue
\mciteSetBstMidEndSepPunct{\mcitedefaultmidpunct}
{\mcitedefaultendpunct}{\mcitedefaultseppunct}\relax
\EndOfBibitem
\bibitem[Marquet \emph{et~al.}(2002)Marquet, Buguin, Talini, and
  Silberzan]{Marquet2002}
C.~Marquet, A.~Buguin, L.~Talini and P.~Silberzan, \emph{Physical Review
  Letters}, 2002, \textbf{88}, 168301\relax
\mciteBstWouldAddEndPuncttrue
\mciteSetBstMidEndSepPunct{\mcitedefaultmidpunct}
{\mcitedefaultendpunct}{\mcitedefaultseppunct}\relax
\EndOfBibitem
\bibitem[Der\'{e}nyi and Vicsek(1995)]{Derenyi1995}
I.~Der\'{e}nyi and T.~Vicsek, \emph{Physical review letters}, 1995,
  \textbf{75}, 374\relax
\mciteBstWouldAddEndPuncttrue
\mciteSetBstMidEndSepPunct{\mcitedefaultmidpunct}
{\mcitedefaultendpunct}{\mcitedefaultseppunct}\relax
\EndOfBibitem
\bibitem[Der\'{e}nyi and Ajdari(1996)]{Derenyi1996}
I.~Der\'{e}nyi and A.~Ajdari, \emph{Physical Review E}, 1996, \textbf{54},
  R5--R8\relax
\mciteBstWouldAddEndPuncttrue
\mciteSetBstMidEndSepPunct{\mcitedefaultmidpunct}
{\mcitedefaultendpunct}{\mcitedefaultseppunct}\relax
\EndOfBibitem
\bibitem[Marathe \emph{et~al.}(2008)Marathe, Jain, and Dhar]{Marathe2008}
R.~Marathe, K.~Jain and A.~Dhar, \emph{J. Stat. Mech. Theory Exp.}, 2008,
  \textbf{2008}, P11014\relax
\mciteBstWouldAddEndPuncttrue
\mciteSetBstMidEndSepPunct{\mcitedefaultmidpunct}
{\mcitedefaultendpunct}{\mcitedefaultseppunct}\relax
\EndOfBibitem
\bibitem[Savel'ev \emph{et~al.}(2004)Savel'ev, Marchesoni, and
  Nori]{Savelev2004}
S.~Savel'ev, F.~Marchesoni and F.~Nori, \emph{Phys. Rev. E}, 2004, \textbf{70},
  061107\relax
\mciteBstWouldAddEndPuncttrue
\mciteSetBstMidEndSepPunct{\mcitedefaultmidpunct}
{\mcitedefaultendpunct}{\mcitedefaultseppunct}\relax
\EndOfBibitem
\bibitem[Pototsky \emph{et~al.}(2010)Pototsky, Archer, Bestehorn, Merkt,
  Savel'ev, and Marchesoni]{Pototsky2010}
A.~Pototsky, A.~J. Archer, M.~Bestehorn, D.~Merkt, S.~Savel'ev and
  F.~Marchesoni, \emph{Phys. Rev. E}, 2010, \textbf{82}, 030401\relax
\mciteBstWouldAddEndPuncttrue
\mciteSetBstMidEndSepPunct{\mcitedefaultmidpunct}
{\mcitedefaultendpunct}{\mcitedefaultseppunct}\relax
\EndOfBibitem
\bibitem[Savel'ev \emph{et~al.}(2003)Savel'ev, Marchesoni, and
  Nori]{Savelev2003}
S.~Savel'ev, F.~Marchesoni and F.~Nori, \emph{Phys. Rev. Lett.}, 2003,
  \textbf{91}, 010601\relax
\mciteBstWouldAddEndPuncttrue
\mciteSetBstMidEndSepPunct{\mcitedefaultmidpunct}
{\mcitedefaultendpunct}{\mcitedefaultseppunct}\relax
\EndOfBibitem
\bibitem[Chakraborty and Chaudhuri(2015)]{Chakraborty2014}
D.~Chakraborty and D.~Chaudhuri, \emph{Physical Review E - Statistical,
  Nonlinear, and Soft Matter Physics}, 2015, \textbf{91}, 050301(R)\relax
\mciteBstWouldAddEndPuncttrue
\mciteSetBstMidEndSepPunct{\mcitedefaultmidpunct}
{\mcitedefaultendpunct}{\mcitedefaultseppunct}\relax
\EndOfBibitem
\bibitem[Chowdhury \emph{et~al.}(1985)Chowdhury, Ackerson, and
  Clark]{Chowdhury1985}
A.~Chowdhury, B.~J. Ackerson and N.~A. Clark, \emph{Phys. Rev. Lett.}, 1985,
  \textbf{55}, 833\relax
\mciteBstWouldAddEndPuncttrue
\mciteSetBstMidEndSepPunct{\mcitedefaultmidpunct}
{\mcitedefaultendpunct}{\mcitedefaultseppunct}\relax
\EndOfBibitem
\bibitem[Wei \emph{et~al.}(1998)Wei, Bechinger, Rudhardt, and
  Leiderer]{Wei1998}
Q.-H. Wei, C.~Bechinger, D.~Rudhardt and P.~Leiderer, \emph{Physical Review
  Letters}, 1998, \textbf{81}, 2606--2609\relax
\mciteBstWouldAddEndPuncttrue
\mciteSetBstMidEndSepPunct{\mcitedefaultmidpunct}
{\mcitedefaultendpunct}{\mcitedefaultseppunct}\relax
\EndOfBibitem
\bibitem[Frey \emph{et~al.}(1999)Frey, Nelson, and Radzihovsky]{Frey1999}
E.~Frey, D.~R. Nelson and L.~Radzihovsky, \emph{Phys. Rev. Lett.}, 1999,
  \textbf{83}, 2977\relax
\mciteBstWouldAddEndPuncttrue
\mciteSetBstMidEndSepPunct{\mcitedefaultmidpunct}
{\mcitedefaultendpunct}{\mcitedefaultseppunct}\relax
\EndOfBibitem
\bibitem[Chaudhuri and Sengupta(2006)]{Chaudhuri2006}
D.~Chaudhuri and S.~Sengupta, \emph{Physical Review E}, 2006, \textbf{73},
  11507\relax
\mciteBstWouldAddEndPuncttrue
\mciteSetBstMidEndSepPunct{\mcitedefaultmidpunct}
{\mcitedefaultendpunct}{\mcitedefaultseppunct}\relax
\EndOfBibitem
\bibitem[Kosterlitz and Thouless(1973)]{Kosterlitz1973}
J.~M. Kosterlitz and D.~J. Thouless, \emph{J. Phys. C}, 1973, \textbf{6},
  1181\relax
\mciteBstWouldAddEndPuncttrue
\mciteSetBstMidEndSepPunct{\mcitedefaultmidpunct}
{\mcitedefaultendpunct}{\mcitedefaultseppunct}\relax
\EndOfBibitem
\bibitem[Halperin and Nelson(1978)]{Halperin1978}
B.~I. Halperin and D.~R. Nelson, \emph{Phys. Rev. Lett.}, 1978, \textbf{41},
  121--124\relax
\mciteBstWouldAddEndPuncttrue
\mciteSetBstMidEndSepPunct{\mcitedefaultmidpunct}
{\mcitedefaultendpunct}{\mcitedefaultseppunct}\relax
\EndOfBibitem
\bibitem[Young(1979)]{Young1979}
A.~P. Young, \emph{Phys. Rev. B}, 1979, \textbf{19}, 1855\relax
\mciteBstWouldAddEndPuncttrue
\mciteSetBstMidEndSepPunct{\mcitedefaultmidpunct}
{\mcitedefaultendpunct}{\mcitedefaultseppunct}\relax
\EndOfBibitem
\bibitem[Kapfer and Krauth(2015)]{Kapfer:2015ca}
S.~C. Kapfer and W.~Krauth, \emph{Physical Review Letters}, 2015, \textbf{114},
  035702--5\relax
\mciteBstWouldAddEndPuncttrue
\mciteSetBstMidEndSepPunct{\mcitedefaultmidpunct}
{\mcitedefaultendpunct}{\mcitedefaultseppunct}\relax
\EndOfBibitem
\bibitem[Frenkel and Smit(2002)]{Frenkel2002}
D.~Frenkel and B.~Smit, \emph{{Understanding molecular simulation: from
  algorithms to applications}}, Academic press, NY, 2002\relax
\mciteBstWouldAddEndPuncttrue
\mciteSetBstMidEndSepPunct{\mcitedefaultmidpunct}
{\mcitedefaultendpunct}{\mcitedefaultseppunct}\relax
\EndOfBibitem
\bibitem[Lahtinen \emph{et~al.}(2001)Lahtinen, Hjelt, Ala-Nissila, and
  Chvoj]{Lahtinen2001}
J.~Lahtinen, T.~Hjelt, T.~Ala-Nissila and Z.~Chvoj, \emph{Physical Review E},
  2001, \textbf{64}, 021204\relax
\mciteBstWouldAddEndPuncttrue
\mciteSetBstMidEndSepPunct{\mcitedefaultmidpunct}
{\mcitedefaultendpunct}{\mcitedefaultseppunct}\relax
\EndOfBibitem
\bibitem[Falck and Lahtinen(2004)]{Falck2004}
E.~Falck and J.~Lahtinen, \emph{The European Physical Journal E}, 2004,
  \textbf{13}, 267\relax
\mciteBstWouldAddEndPuncttrue
\mciteSetBstMidEndSepPunct{\mcitedefaultmidpunct}
{\mcitedefaultendpunct}{\mcitedefaultseppunct}\relax
\EndOfBibitem
\bibitem[Chaikin and Lubensky(2012)]{Chaikin2012}
P.~M. Chaikin and T.~C. Lubensky, \emph{Principles of Condensed Matter
  Physics}, Cambridge University Press, Cambridge, 2012\relax
\mciteBstWouldAddEndPuncttrue
\mciteSetBstMidEndSepPunct{\mcitedefaultmidpunct}
{\mcitedefaultendpunct}{\mcitedefaultseppunct}\relax
\EndOfBibitem
\bibitem[Mickel \emph{et~al.}(2013)Mickel, Kapfer, Schr{\"{o}}der-Turk, and
  Mecke]{Mickel2013}
W.~Mickel, S.~C. Kapfer, G.~E. Schr{\"{o}}der-Turk and K.~Mecke, \emph{J. Chem.
  Phys.}, 2013, \textbf{138}, 044501\relax
\mciteBstWouldAddEndPuncttrue
\mciteSetBstMidEndSepPunct{\mcitedefaultmidpunct}
{\mcitedefaultendpunct}{\mcitedefaultseppunct}\relax
\EndOfBibitem
\bibitem[Qi \emph{et~al.}(2014)Qi, Gantapara, and Dijkstra]{Qi:2014}
W.~Qi, A.~P. Gantapara and M.~Dijkstra, \emph{Soft Matter}, 2014, \textbf{10},
  5449--5457\relax
\mciteBstWouldAddEndPuncttrue
\mciteSetBstMidEndSepPunct{\mcitedefaultmidpunct}
{\mcitedefaultendpunct}{\mcitedefaultseppunct}\relax
\EndOfBibitem
\end{mcitethebibliography}
\end{document}